%% file: bagley_ms.tex
\documentclass[trackchanges,twocolumn]{aastex62}
\usepackage{natbib}
\usepackage{url}
\usepackage{color}
\usepackage[super]{nth}
\input{definitions}

\accepted{May 30, 2020}
\submitjournal{ApJ}

\shorttitle{\textit{HST} Grism-derived Forecasts}
\shortauthors{Bagley et al.}

\begin{document}

\title{\textit{HST} Grism-derived Forecasts for Future Galaxy Redshift Surveys}

\correspondingauthor{Micaela B. Bagley}
\email{mbagley@utexas.edu}

\author[0000-0002-9921-9218]{Micaela B. Bagley}
\affil{University of Texas at Austin, 
2515 Speedway, Stop C1400, 
Austin, Texas 78712-1205, USA}
\affil{Minnesota Institute for Astrophysics,
University of Minnesota,
Minneapolis, MN 55455, USA}
\author{Claudia Scarlata}
\affil{Minnesota Institute for Astrophysics,
University of Minnesota,
Minneapolis, MN 55455, USA}
\author{Vihang Mehta}
\affil{Minnesota Institute for Astrophysics,
University of Minnesota,
Minneapolis, MN 55455, USA}
\author{Harry Teplitz}
\affil{Infrared Processing and Analysis Center, 
California Institute of Technology, 
Pasadena, CA 91125, USA}
\author{Ivano Baronchelli}
\affil{Dipartimento di Fisica e Astronomia ``G. Galilei'', 
Universit\`{a} degli Studi di Padova, 
Vicolo dell'Osservatorio 3, I-35122, Italy}
\affil{Infrared Processing and Analysis Center,
California Institute of Technology,
Pasadena, CA 91125, USA}
\author{Daniel J. Eisenstein}
\affil{Harvard-Smithsonian Center for Astrophysics, 
60 Garden St., 
Cambridge, MA 02138, USA}
\author{Lucia Pozzetti}
\affil{INAF -- Osservatorio di Astrofisica e Scienza dello Spazio
di Bologna, Bologna, I-40129, Italy}
\author{Andrea Cimatti}
\affil{University of Bologna, 
Department of Physics and Astronomy,
Via Gobetti 93/2, I-40129, Bologna, Italy}
\affil{INAF -- Osservatorio Astrofisico di Arcetri, 
Largo E. Fermi 5, I-50125, Firenze, Italy}
\author{Michael Rutkowski}
\affil{Department of Physics and Astronomy, 
Minnesota State University, 
Mankato, MN 56001, USA}
\author{Yun Wang}
\affil{Infrared Processing and Analysis Center,
California Institute of Technology,
Pasadena, CA 91125, USA}
\author{Alexander Merson}
\affil{Jet Propulsion Laboratory, 
California Institute of Technology, 
4800 Oak Grove Drive, 
Pasadena, CA 91109, USA}
\affil{Infrared Processing and Analysis Center,
California Institute of Technology,
Pasadena, CA 91125, USA}

%%%%%%%%%%%%%%%%%%%%%%%%%%%%%%%%%%%%%%%%
\begin{abstract}
The mutually complementary Euclid and \textit{Roman} galaxy redshift 
surveys will use \ha- and \oiii-selected emission line galaxies as tracers of 
the large scale structure at $0.9 \lesssim z \lesssim 1.9$ (\ha) and 
$1.5 \lesssim z \lesssim 2.7$ (\oiii). 
It is essential to have a reliable and sufficiently 
precise knowledge of the expected numbers of \ha-emitting galaxies in the 
survey volume in order to optimize these redshift surveys for 
the study of dark energy.
Additionally, these future samples of emission-line galaxies 
will, like all slitless spectroscopy surveys, be affected by a complex 
selection function that depends on galaxy size and luminosity, line equivalent 
width, and redshift errors arising from the misidentification of single 
emission-line galaxies. 
Focusing on the specifics of the Euclid survey, we combine two 
slitless spectroscopic WFC3-IR datasets -- \tdhst\ and the WISP survey -- to 
construct a Euclid-like sample that covers an area of 0.56~deg$^2$ and 
includes 1277 emission line galaxies. 
We detect 1091 ($\sim$3270 deg$^{-2}$) \han-emitting galaxies in the 
range $0.9\leq z \leq 1.6$ and 162 ($\sim$440 deg$^{-2}$) \oiiib-emitters over 
$1.5\leq z \leq 2.3$ with line fluxes $\geq2\times 10^{-16}$ \esc. The median of the \han\
equivalent width distribution is $\sim$250~\AA, and the 
effective radii of the continuum and \han\ emission are correlated with a 
median of $\sim0\farcs38$ and
significant scatter ($\sigma \sim 0\farcs2-0\farcs35$).
Finally, we explore the prevalence of redshift mis-identification in future 
Euclid samples, finding potential contamination rates of $\sim$14-20\% 
and $\sim$6\% down to $2\times 10^{-16}$ and $6 \times 10^{-17}$ \esc, 
respectively, though with increased wavelength coverage these percentages
drop to nearly zero.
\end{abstract}

\keywords{Emission line galaxies (459); Redshift surveys (1378); 
Spectroscopy (1558)}

%%%%%%%%%%%%%%%%%%%%%%%%%%%%%%%%%%%%%%%%
\section{Introduction} \label{sec:intro}
The nature of dark energy, the explanation of the observed cosmic 
acceleration \citep{riess1998,perlmutter1999}, is one of the most important 
unsolved problems in cosmology today. A galaxy redshift survey enables us 
to measure the cosmic expansion history via the measurement of baryon acoustic 
oscillations (BAO), as well as the growth history of large scale structure 
via the measurement of large scale redshift-space distortions. The combination 
of these two measurements allows us to differentiate between an unknown energy 
component and the modification of general relativity as the cause of the 
observed cosmic acceleration \citep{guzzo2008, wang2008a}.

Two future space missions, ESA's Euclid \citep{laureijs2011, laureijs2012} 
and NASA's \textit{Nancy Grace Roman Space Telescope} 
\citep[\rst, formerly WFIRST;][]{green2012, spergel2015}, 
will carry out mutually 
complementary galaxy redshift surveys to probe dark energy. Both Euclid and 
\rst\ will use \ha\ and \oiii-selected emission line galaxies as tracers of 
the large scale structure  at $0.9 \lesssim z \lesssim 1.9$ (\ha) and 
$1.5\lesssim z \lesssim 2.7$ (\oiii). The uncertainties in the cosmological 
parameters derived from a BAO survey are inversely proportional to the number 
of galaxies used in the survey. To optimize these redshift surveys for the 
study of dark energy, it is therefore critical to have a reliable and 
sufficiently precise knowledge of the expected numbers of \ha\ and \oiii\ 
galaxies in the survey volume.

In the redshift range of interest for the galaxy redshift surveys ($0.9<z<2.7$),
existing \ha\ and \oiii\ luminosity function measurements show large
uncertainties and are often inconsistent with one another. In the relevant 
redshift range, \ha\ and \oiii-emitting galaxies are identified with two main 
techniques. Ground-based narrowband surveys 
\citep[e.g., ][]{geach2008,sobral2009b} cover large areas, but are limited by 
very thin redshift slices ($\Delta (z)\sim 0.03$). Slitless space-based 
spectroscopic surveys, with NICMOS first \citep[e.g., ][]{hopkins2000,shim2009}
and WFC3 more recently \citep{colbert2013,mehta2015,pirzkal2017}, 
simultaneously probe a large redshift range ($\Delta (z) \sim 0.7$), 
albeit over much smaller areas. Despite the enormous effort, the uncertainties 
on the luminosity functions remain substantial. For example, the characteristic luminosities, 
$L^*$, measured from a variety of surveys across this redshift range span 
almost an order of magnitude \citep[e.g., ][]{hopkins2000,geach2008,hayes2010,colbert2013,sobral2013,mehta2015,matthee2017}.
These uncertainties lead to less certain number count predictions for
galaxy redshift surveys such as those of Euclid and \rst, 
measurements necessary to constrain dark energy.

In addition to accurate number counts, simulations are also an 
important component of the preparation required for surveys such as these.
Cosmological N-body simulations, hydrodynamical codes, semi-analytic models,    and the mock catalogs generated from them are valuable tools in preparing 
to physically interpret the wealth of measurements that are expected.
Additionally, such models and catalogs can be used to test reduction,
sample selection, and source characterization software being developed
to process and analyze the survey data. In both cases, it
is crucial that these simulations reproduce the
observed joint distributions of emission line fluxes and 
galaxy size, luminosity, and mass, and correctly assign line fluxes as a 
function of these properties. The proper assignment of
galaxy properties is necessary to correctly account for observational
selection effects, which depend on galaxy size and luminosity as well as 
emission line signal-to-noise (S/N) and equivalent width (EW).

There have been significant recent efforts to prepare for future 
galaxy redshift surveys. 
For example, \cite{pozzetti2016} and \cite{merson2018} use 
physically-motivated models to predict the expected number of \ha-emitting 
galaxies that will be detectable down to a range of survey flux limits. 
\cite{valentino2017} similarly predict emission line number counts using 
large, spectroscopically-calibrated photometric samples. 
Others have addressed the important challenges of automatically 
identifying emission lines in slitless data \citep[e.g.,][]{maseda2018} and 
of quantifying the quality of spectroscopic redshifts 
\citep[e.g.,][]{jamal2018}. 
Yet much of this work either makes use of slit-based spectroscopy that has a
distinct selection function from that of slitless data, or requires auxiliary 
datasets such as multi-wavelength photometry. 
In this paper, we add to these works by leveraging the similarities of future 
slitless grisms with those of the \textit{Hubble Space Telescope} (\hst) WFC3 
G102 and G141 infrared grisms to create a selection function 
that closely approximates that of the upcoming galaxy redshift surveys. 
We expand on the work presented in \cite{colbert2013} and \cite{mehta2015},
which has previously been compared with models by 
\citet{pozzetti2016} and \citet{merson2018},
by combining multiple \hst\ grism programs to cover a $>$10$\times$ greater 
area. 
As our results do not depend on photometrically-determined redshifts, 
our work is complementary to that of \cite{maseda2018}. 
The large survey footprints planned for future galaxy redshift surveys will not 
be fully covered by the same wealth of multi-wavelength imaging observations 
that is available
for CANDELS fields, where photometric redshifts are 
based on 8 \citep[UDS;][]{williams2009} to 35 
\citep[COSMOS;][]{whitaker2011} photometric measurements. 
Our results are therefore an important 
representation of the expectations for grism surveys, even for fields 
that will lack the coverage in additional auxiliary imaging datasets 
to obtain sufficiently accurate and precise photometric redshifts.

While the details of the \rst\ survey are still under development,
the Euclid Consortium is in the process of finalizing the observing strategy
for the Euclid mission. In this paper, we therefore focus on the projected
characteristics for Euclid and use available slitless spectroscopic
data from \hst\ grism surveys to make predictions for this survey. 
In what follows we calculate the number densities of \ha\ and \oiii-emitting 
galaxies, measure the size and EW distributions for \ha-emitters,
and quantify the expected number of contaminating redshifts from 
misidentified single emission lines as a function of survey depth and 
redshift. We estimate the number
density of \ha-emitters accessible to galaxy redshift surveys by applying
selection criteria matching those of the Euclid Wide Survey. The Wide Survey 
will use the Near Infrared Spectrograph and Photometer (NISP) to 
detect emission line galaxies (ELGs) in a 15000 deg$^2$ survey area
down to a 3.5$\sigma$ flux limit of $2\times10^{-16}$ \esc\ 
for sources $0\farcs5$ in diameter \citep{racca2016,vavrek2016}. 
We note, however, that similar predictions can be tuned for \rst\ by 
adjusting the selection criteria appropriately.

This paper is organized as follows. 
In Section~\ref{sec:sample}, we present the slitless grism survey 
characteristics and describe the creation of a Euclid-like Wide Sample using 
the Euclid selection function. In preparation for characterizing the 
emission size distributions of the sample, we describe the creation of an 
empirical PSF and our method for fitting models to emission maps in 
Section~\ref{sec:emissionsizes}.
We present our results in Section~\ref{sec:results}, including the 
number counts of emission line galaxies (Section~\ref{sec:counts}),
the continuum and emission line sizes (Section~\ref{sec:sizes}), the 
EW distribution for \han-emitters (Section~\ref{sec:ew}),
and a potential \oiii\ selection bias based on the \oiii\ line profiles 
in the grism data (Section~\ref{sec:oiii}).
We present an empirical measurement of the redshift accuracy achievable with 
slitless grism data in Section~\ref{sec:zaccuracy} and discuss the effects of
contamination from misidentified single emission lines in 
Section~\ref{sec:contam}. 
Finally, we summarize the key results in Section~\ref{sec:summary}.
Throughout this paper we assume a $\Lambda$CDM cosmology with $\Omega_M=0.3$, 
$\Omega_{\Lambda}=0.7$, and $H_0 = 70$ km s$^{-1}$ Mpc$^{-1}$. All 
magnitudes are expressed in the AB system \citep{oke1983}.

%%%%%%%%%%%%%%%%%%%%%%%%%%%%%%%%%%%%%%%%
\section{The Euclid-like Sample} \label{sec:sample}
\begin{figure}
\epsscale{1.2}
\plotone{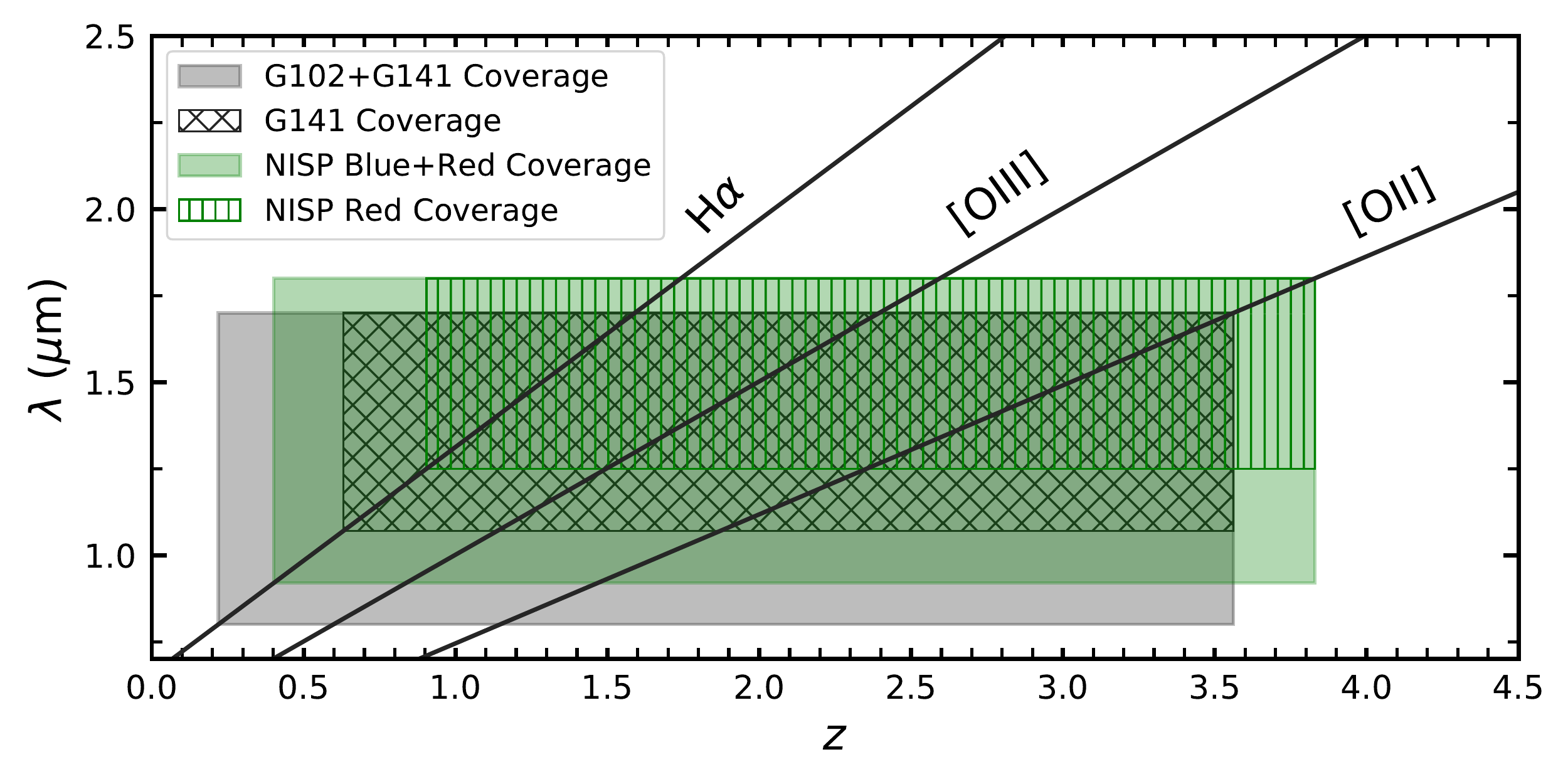}
\caption{
The redshift and wavelength coverage of the WFC3 grisms (gray) is compared 
with that of the Euclid NISP grisms (green). 
The wavelengths of \ha, \oiii, and \oii\ as a function of redshift are 
designated by black lines. The shaded and hatched regions indicate the
redshift range in which at least one of these three emission lines is 
accessible to the given grism. The coverage of the WFC3 grisms is 
comparable to that planned for Euclid, making the HST grisms 
important tools for exploring the performance of the upcoming galaxy 
redshift survey.
\label{fig:zrange}}
\end{figure}

For this work, we use existing spectroscopic data from three \textit{HST} grism
programs: the WFC3 Infrared Spectroscopic Parallel survey 
\citep[WISP, see Section~\ref{sec:wisp};][]{atek2010},
3D-HST \citep[Section~\ref{sec:3dhst};][]{brammer2012,skelton2014,momcheva2016},
and A Grism H-Alpha SpecTroscopic survey 
\citep[AGHAST, Section~\ref{sec:3dhst};][]{weiner2009}.
All programs perform near-infrared slitless spectroscopic observations 
using one or both of the WFC3\footnote{\url{www.stsci.edu/hst/wfc3}} 
IR grisms: G102 ($0.8-1.1 \mu$m, $R\sim210$) and 
G141 ($1.07-1.7 \mu$m, $R\sim130$).
The wavelength range of the G141 grism in particular covers a comparable
redshift range as that planned by the Euclid galaxy redshift survey
(see Figure~\ref{fig:zrange}). The IR channel of the WFC3 \citep{kimble2008} 
has a field of view of $123\arcsec \times 134\arcsec$ and a native pixel 
scale of $0\farcs13$/pixel. The WFC3 observations compiled here from the 
WISP, 3D-HST, and AGHAST surveys cover a total area of 0.56 deg$^2$,
which is approximately equal to the NISP field of view.
The sources detected by these three surveys, while not necessarily 
representative of the full population of galaxies,
are representative of the galaxies accessible to similar grism surveys.

%%%%%%%%%%%%%%%%%%%%%%%%%%%%%%%%%%%%%%%%
\subsection{The WISP Survey} \label{sec:wisp}
The WISP Survey \citep[PI: M. Malkan;][]{atek2010} is an \hst\ pure 
parallel program,
obtaining WFC3 observations of nearby fields while other \hst\ instruments 
are in use. In particular, WISP observations are taken in parallel when 
either the Cosmic Origins Spectrograph \citep[COS; ][]{froning2009} or the 
Space Telescope Imager and Spectrograph \citep[STIS; ][]{kimble1998} are 
used as the primary instrument, as programs with these two instruments 
typically involve long integrations of a single pointing. The WISP parallel 
field is offset by $\sim5\arcmin$ from the primary target. Since the 
selection of parallel opportunities depends on the integration time rather 
than the position of the primary target, WISP fields are independent and 
uncorrelated. 
In this paper, we include emission line measurements from 419 WISP fields
collectively covering $\sim$1520 arcmin$^2$.

The WISP observing strategy depends on the details of each parallel 
opportunity, and therefore varies from field to field. In short opportunities 
consisting of one to three continuous orbits, the G141 grism is typically used 
along with one imaging filter (F140W or F160W) to aid in spectral extraction 
and to mark the zero point for wavelength calibration. The G102 grism and the 
F110W imaging filter are added to longer opportunities consisting of four 
or more continuous orbits. For these deeper fields, the integration times 
in the two grisms are tuned to achieve approximately uniform sensitivity 
for an emission line of a given flux across the full wavelength range. 
As the visit lengths depend on the specifics of the primary observations, 
we do not reach a uniform depth in all WISP fields.
Additionally, the sky background in each field is affected to varying 
degrees by, for example, zodiacal light and Earth limb brightening. 
The median $5\sigma$ detection limit for emission lines in both grisms is
$\sim$5$\eseven$ \esc, yet the detection limit in 
a given field can differ from this median by more than a factor of 2. As a 
consequence, while all WISP fields are deeper than the Euclid 
Wide Survey, only $\sim$75\% of fields reach the expected depth of the Euclid
Deep Survey.

All WISP data are reduced with the WFC3 pipeline CALWF3 in combination with 
custom scripts that account for the specific challenges of un-dithered, 
pure-parallel observations. The foundation of the WISP reduction pipeline
is described in \cite{atek2010}, and crucial updates implemented for the 
current version will be presented in Baronchelli et al. 
(\textit{in prep}).
We use the \texttt{AstroDrizzle} software \citep{drizzlepac} to combine the 
individual exposures, correcting for astrometric distortions and any 
potential alignment issues. The IR direct images are drizzled onto 
a $0\farcs08$/pixel scale. Object detection in the IR direct images 
(F110W, F140W, and F160W) is performed with \se\ 
\citep[version 2.5; ][]{bertin1996}. For fields with imaging in two filters, 
we create a combined detection image and supplement the catalog with sources
detected individually in only one of the filters. 
We use the \texttt{aXe} software package \citep{kummel2009} to extract and 
calibrate the spectra. 
The \texttt{aXe} software drizzles all extracted spectral stamps 
from individual exposures to a combined spectral image with a constant 
dispersion and cross-dispersion pixel scale, thus removing geometric 
distortions.
The individual drizzled spectral stamps are on the $0\farcs13$/pixel scale. 
For each source identified in the direct imaging, the spatial width of 
the extraction window is a factor of $4\times$ the projected size of the 
source (either semi-major or semi-minor axis 
depending on the source orientation) onto the extraction 
direction\footnote{See Figure 1.12 of the \texttt{aXe} User Manual 
(version 2.3),
\url{www.stsci.edu/institute/software_hardware/stsdas/axe/extract_calibrate/axe_manual}}. We then use \texttt{aXe}'s optimal weighting method with Gaussian 
weights (with widths based on the size of the sources in the direct image) 
to extract 1D spectra from the 2D spectral stamps. 
The emission line finding process described in the next section is performed 
on the 1D spectra.

%%%%%%%%%%%%%%%%%%%%%%%%%%%%%%%%%%%%%%%%
\subsubsection{WISP Emission Line Catalog} \label{sec:linecatalog}
We construct the WISP emission line catalog via the combination of an 
automatic detection algorithm that identifies emission line candidates 
and a visual inspection of each candidate performed by two reviewers.
There are two versions of the WISP emission line detection algorithm.
The first version, presented in \cite{colbert2013} for $\sim$30 WISP fields, 
identified emission lines as groups of contiguous pixels about the continuum. 
The resulting lists of line candidates identified by the detection 
algorithm were dominated by spurious sources and 
fake emission lines, and the inspection and cleaning of these lists required 
extensive time and effort from reviewers. We developed the new method 
to substantially reduce the time required for reviewers to inspect all 
400$+$ WISP fields by improving the methods for automatic identification
and vetting of emission line candidates. This second version of the 
detection algorithm improves on this method by including a continuous 
wavelet transform, which fits not only the amplitude, but also the shape, of
emission line features in a spectrum. The new algorithm also includes 
additional quality checks aiming to remove most spurious sources before 
the inspection stage. The details of the new algorithm will be presented in 
an upcoming paper, Bagley et al. (\textit{in prep}).

Following detection, each emission line candidate is visually inspected by two 
reviewers to reject artifacts such as cosmic rays and hot pixels, to remove 
lines that are heavily contaminated by overlapping spectra, and to identify 
the emission lines and fit the source redshift. 
The full spectrum is then fit with a single model consisting of 
a continuum and Gaussian emission lines at wavelengths determined by the 
redshift assigned to the source.
Specifically, the reviewer provides an initial guess at the source 
redshift by identifying an emission feature. The best-fit redshift is 
determined using a least-squares minimization of the full emission model, 
including emission lines and the continuum. The redshift of the fit is 
constrained to be between $\Delta z \pm 0.02$ of the initial guess, which 
corresponds to $\sim$130~\AA, or $\sim$$\pm$3 pixels at the dispersion of the 
G141 grism. The peak wavelength of each additional emission line in the 
spectrum is allowed to vary by the same amount to allow for any offsets from 
the systemic redshift and/or centering differences due to the low resolution 
of the spectra.
% Black Lives Matter

This method ensures that all emission lines are fit with profiles of the 
same full width at half maximum (FWHM)\footnote{
For each source, the best-fit FWHM is determined by the emission 
model fitting and the initial guess depends on the source size in the 
direct imaging as follows.
The semi-major axis ($a$) is used as an approximate FWHM in 
pixels and multiplied by the 
grism dispersion ($\Delta\lambda$):
FWHM$_{\mathrm{init}}=2.35 a \ \mathrm{[pixel]} \ \Delta\lambda \ [\mathrm{\AA/pixel}]$.}, appropriate for slitless spectra 
where all emission lines are images of the same host source. 
Simultaneously fitting all emission lines 
also helps eliminate contamination from overlapping spectra, as the 
wavelengths of lines from other sources will not match the model for the 
given source redshift. As a consequence of this simultaneous fitting, 
fluxes or upper limits are measured for all lines in the wavelength range 
determined by the assigned redshift, whether or not the lines were identified 
by the detection algorithm. The WISP emission line catalog therefore 
contains both ``primary'' emission lines detected by the automatic peak 
finder and ``secondary'' lines that often have a lower S/N than the 
detection threshold.  
This distinction is relevant for the application of completeness 
corrections (see below) and can have important implications for sample 
selection. While sources in the emission line catalog can have multiple 
primary lines (usually \ha\ and \oiii)
secondary lines (often \siii~$\lambda\lambda9069,~9532$ for example)
are measured as a consequence of a primary line detection.
Finally, we note that in the absence of multiple emission lines, 
single lines are assumed to be \ha\ unless the clear asymmetry of the 
\oiii$+$\hb\ line profile is visible. We discuss this assumption further in 
Section~\ref{sec:oiii}.

The WISP emission line catalog was constructed after processing and 
inspecting the spectra from 419 WISP fields, covers $\sim$1520 arcmin$^2$,
and includes $\sim$8000 emission line objects.
The improved emission line detection process and completeness analysis 
will be presented in Bagley et al. (\textit{in prep}), and the 
resulting emission line catalog will be released at the time of publication.
We use this catalog, in combination with that from \tdhst\ discussed in 
Section~\ref{sec:3dhst}, to construct a Euclid-like sample in 
Section~\ref{sec:selection}.

%%%%%%%%%%%%%%%%%%%%%%%%%%%%%%%%%%%%%%%%
\subsection{The \tdhst\ Survey} \label{sec:3dhst}
The 3D-HST Survey \citep[PI: P. van Dokkum;][]{brammer2012,skelton2014,momcheva2016}
and the AGHAST Survey \citep[PI: B. Weiner;][]{weiner2009} 
together obtained spectroscopic
observations of the CANDELS \citep{grogin2011,koekemoer2011} fields.
In $\sim$150 pointings, the \tdhst\ Survey covered each field to a uniform
two-orbit depth, including G141 observations and direct imaging in the F140W
filter.
We add the \tdhst\ pointings from the AEGIS, COSMOS, GOODS-North (GOODS-N) and
GOODS-South (GOODS-S) fields, $\sim$507 arcmin$^2$ in total, to the WISP fields.
Including these fields in our analysis has several benefits in addition to
the increase in area coverage. With the extensive multi-wavelength
catalogs available for the well-studied CANDELS fields, we can identify
regions in color space indicative of misidentified single emission lines
(see Section~\ref{sec:contam}). 

\begin{deluxetable*}{r|ll}
\centering
\tablewidth{0pt}
\tablecaption{Euclid Sample Selection Criteria \label{tab:selection}}
\tablehead{
\hspace{1cm} & \textbf{Euclid WS} &  \textbf{Euclid DS}}
%\colhead{} & \colhead{Euclid WS} & \colhead{Euclid DS}}
\startdata
S/N & $\geq5$ & $\geq5$ \\
EW$_{\mathrm{obs}}$ & $\geq40$~\AA\ & $\geq40$~\AA\ \\
Flux & $\geq 2\times 10^{-16}$ \esc\ & $\geq 6\times 10^{-17}$ \esc\ \\
$\lambda_{\mathrm{obs}}$ & $\geq 12500$~\AA\ & $\geq 9200$~\AA\ \\[5pt]
\hline \\[-7pt]
\ha\ Coverage & $0.9 \leq z \leq 1.6$ & $0.4 \leq z \leq 1.6$ \\
\oiii\ Coverage & $1.5 \leq z \leq 2.4$ & $0.8 \leq z \leq 2.4$ \\
\oii\ Coverage & $2.5 \leq z \leq 3.5$ & $1.5 \leq z \leq 3.5$ \\
\enddata
\tablecomments{In this paper we focus on observational constraints relevant for 
the Euclid Wide Survey and create a Wide Sample (WS) with these criteria. The
Euclid DS selection criteria are presented here for reference. 
In Section~\ref{sec:contam} we extend our 
analysis of the WISP$+$3D-HST catalog down to the expected flux and 
wavelength limits 
of the Deep Survey in order to explore sample contamination from redshift
misidentification.}
\end{deluxetable*}

The \tdhst\ team has released a catalog with emission line measurements for 
all galaxies detected in imaging \citep{momcheva2016}. Their method involves
combining the CANDELS photometry with the grism spectroscopy to determine 
augmented photometric redshifts, which are then used as a prior for detecting 
and measuring emission lines in the grism data. 
The Euclid Wide survey observations will, at a minimum, include imaging in the
$Y$, $J$, and $H$ filters of the NISP instrument as well as the
very broad VIS filter covering $\sim5500-9000$~\AA. While additional
ground-based imaging in the $g$, $r$, $i$, and $z$ bands will be obtained,
the SEDs of sources will not be as fully sampled as in the CANDELS fields.
The amount of information Euclid obtains for each source will be closer to 
the level obtained in WISP observations. We therefore reprocess all 
\tdhst\ data in a consistent manner with the WISP fields \citep{rutkowski2016}.
We note, however, that time scales and effort required to run even the 
improved WISP emission line procedure on all 15,000 deg$^2$ of the Euclid 
Wide survey will be impossibly unrealistic.
Alternative emission line detection algorithms will be needed, such as the 
citizen science pilot program described in \cite{dickinson2018} 
or the integration of machine learning and human classification 
such as that of \cite{beck2018}

\cite{rutkowski2016} describe the reduction of the \tdhst\ data using the 
WISP pipeline with minor modifications to account for the dithered 
observations as well as the creation of the \tdhst\ emission line catalog. 
Emission line detection and measurement are performed 
using the first version of the WISP line finding procedure, which is 
presented in \cite{colbert2013} and discussed in 
Appendix~\ref{app:tdhstcorrections}. Briefly, emission line candidates are 
identified as groups of contiguous pixels above the continuum. 
In contrast to the WISP catalog (Section~\ref{sec:linecatalog}), we 
fit each \tdhst\ emission line individually, therefore measuring the 
redshift, flux, FWHM, and EW separately for each line. 
Single, symmetric emission lines are again assumed to be \ha.
The \tdhst\ catalog 
includes $\sim$5700 emission line objects, and is combined with the WISP 
catalog in Section~\ref{sec:selection}.

%%%%%%%%%%%%%%%%%%%%%%%%%%%%%%%%%%%%%%%%
\subsection{Emission line catalog completeness corrections}\label{sec:completeness}

Grism surveys such as WISP, \tdhst, and Euclid can suffer from
incompleteness for a variety of reasons. Sources may be lost amidst the
noise in images if their fluxes are close to the detection limit. Some
sources may not be detected, or their emission lines missed in their
spectra, because they overlap or are blended with nearby bright objects.
The completeness of a survey depends on the specific selection function
used to detect sources. In the case of the WISP and 3D-HST$+$AGHAST emission 
line catalogs, the selection function includes the detection of the sources 
in the direct images, the identification of emission line candidates via the
detection algorithm, and the acceptance during visual inspection.

The completeness corrections applied to the WISP and \tdhst\ emission 
line catalogs were derived in a manner consistent with each of the emission 
line detection procedures. These derivations are similar but not identical for 
the two catalogs, reflecting the differences in the line finding algorithms, 
visual inspection, and emission line fitting.
Specifically, the completeness corrections 
from \citet{colbert2013} are adopted for the \tdhst\ catalog, while a new 
set of simulations is used to determine the completeness of the updated 
line finding procedure that created the WISP catalog. Each method is 
described in more detail in Appendix~\ref{app:completeness}.

Finally, in the sample selection presented in the following section, we have 
adopted additional selection criteria: line EW$_{\mathrm{obs}} > 40$~\AA\ 
and S/N$>$5. 
We discuss the motivation behind these two additional criteria in 
Appendix~\ref{app:completeness}.
We note, however, that while these two criteria are applicable to the
emission line detection processes used for both the WISP and \tdhst\ datasets
in this paper, they will not necessarily be appropriate for Euclid or other
future grism surveys.

%%%%%%%%%%%%%%%%%%%%%%%%%%%%%%%%%%%%%%%%
\subsection{Sample Selection}\label{sec:selection}
The Euclid Mission will be composed of two surveys. The Wide Survey aims 
to obtain redshift measurements for $\sim$25 million galaxies over 15000 
deg$^2$ \citep[e.g.,][]{vavrek2016}, using the Euclid Red grism
($1.25 - 1.85 \mu$m, $R\sim 380$) and achieving a $3.5\sigma$ line flux
sensitivity of $2 \times 10^{-16}$ \esc\ for a source with a 
diameter\footnote{
As emission line fluxes obtained through slitless 
spectroscopy depend on source size, a single flux limit is not fully 
representative of what the Wide Survey will detect. 
More compact sources 
may be detectable down to fainter fluxes, and the distribution of sources 
increases rapidly toward fainter emission line fluxes. However, following 
the example of \cite{laureijs2011}, we adopt here a single flux limit for all 
sources, noting that our analysis therefore represents a conservative 
estimate of the number density of sources available to Euclid.}
of $0\farcs5$.
The Deep Survey will cover 40 deg$^2$ in three separate pointings,
reaching a depth of $6 \times 10^{-17}$ \esc. In addition to the Red grism,
the Deep Survey may make use of a Blue grism ($0.92 - 1.3 \mu$m, 
with a tentative $R\sim250$).
In this paper, we focus on observational constraints relevant for the 
Wide Survey but note that \hst\ grism observations are valuable for 
Deep Survey predictions as well. 

From the full WISP$+$3D-HST catalog, we create a ``Wide Sample'' (WS) of ELGs
selected to match the planned Euclid Wide Survey.
We leave the construction of a ``Deep Sample'' (DS) for future work, but  
capitalize on the depth of the WISP$+$3D-HST catalog to discuss 
contamination and redshift misidentification in 
Section~\ref{sec:contam}.
We begin by considering only sources with secure redshifts, where either both 
reviewers agree on the assigned redshift or multiple, high-S/N lines are
detected in the source's spectrum. Next we impose a selection
in emission line S/N and observed EW to match the
completeness limits of the full WISP$+$3D-HST catalog: S/N$>5$ and
EW$_{\mathrm{obs}} \geq 40$~\AA\ \citep[see Appendix~\ref{app:completeness} 
as well as ][]{colbert2013}. 
For galaxies at $z\sim1-1.5$, EW$_{\mathrm{obs}} > 40$~\AA\ corresponds 
to a rest EW of $\sim 16-20$~\AA.
The remaining selection criteria depend on emission line flux and
observed wavelength. For the WS, we select sources with at least one emission
line with flux $f \geq 2\times 10^{-16}$ \esc\ and
$\lambda_{\mathrm{obs}} \geq 1.25\mu$m. 
The DS will include additional sources
down to $f \geq 6 \times 10^{-17}$ \esc\ and, in fields observed with the 
Blue grism, 
$\lambda_{\mathrm{obs}} \geq 0.92\mu$m. Given the drop in the sensitivity
of the G141 grism at wavelengths longer than $\sim1.7 \mu$m,
this wavelength selection results in \ha\ (\oiii) coverage from
$0.9 \leq z \leq 1.6$ ($1.5 \leq z \leq 2.4$) for the WS and
$0.4 \leq z \leq 1.6$ ($0.8 \leq z \leq 2.4$) for the DS, respectively.
See Table~\ref{tab:selection} for a summary of the selection criteria.

We note that given the spectral resolution of the planned missions, \ha$+$\nii\
will be blended for most sources in Euclid (and some sources in \rst) spectra.
These two emission lines are also blended in observations
obtained with the WFC3 grisms. For the
purpose of predicting the number, size and EW distributions of the
\ha-emitters that will be detected by the galaxy redshift surveys, we do not
correct the observed \ha\ fluxes for the contribution by \nii.
All measurements presented here of \ha\ flux, EW, and size
refer to \han. 

Similarly, the \oiii~$\lambda\lambda4959,5007$ doublet is partially blended
at the resolution of the WFC3 grisms. The \oiii\ fluxes are obtained by fitting
two blended Gaussians of the same FWHM to the doublet line profile 
using amplitudes fixed in a $1:3$ ratio, following the theoretical 
calculations of \cite{storey2000}. 
Since the \oiii\ doublet will be resolved by Euclid and
\rst, we correct the observed \oiiib\ flux for the contribution
from the 4959\AA\ line using the same flux ratio.
All measurements presented here of \oiii\ flux
therefore refer to \oiiib\ only.
\begin{figure}
\epsscale{1.1}
\plotone{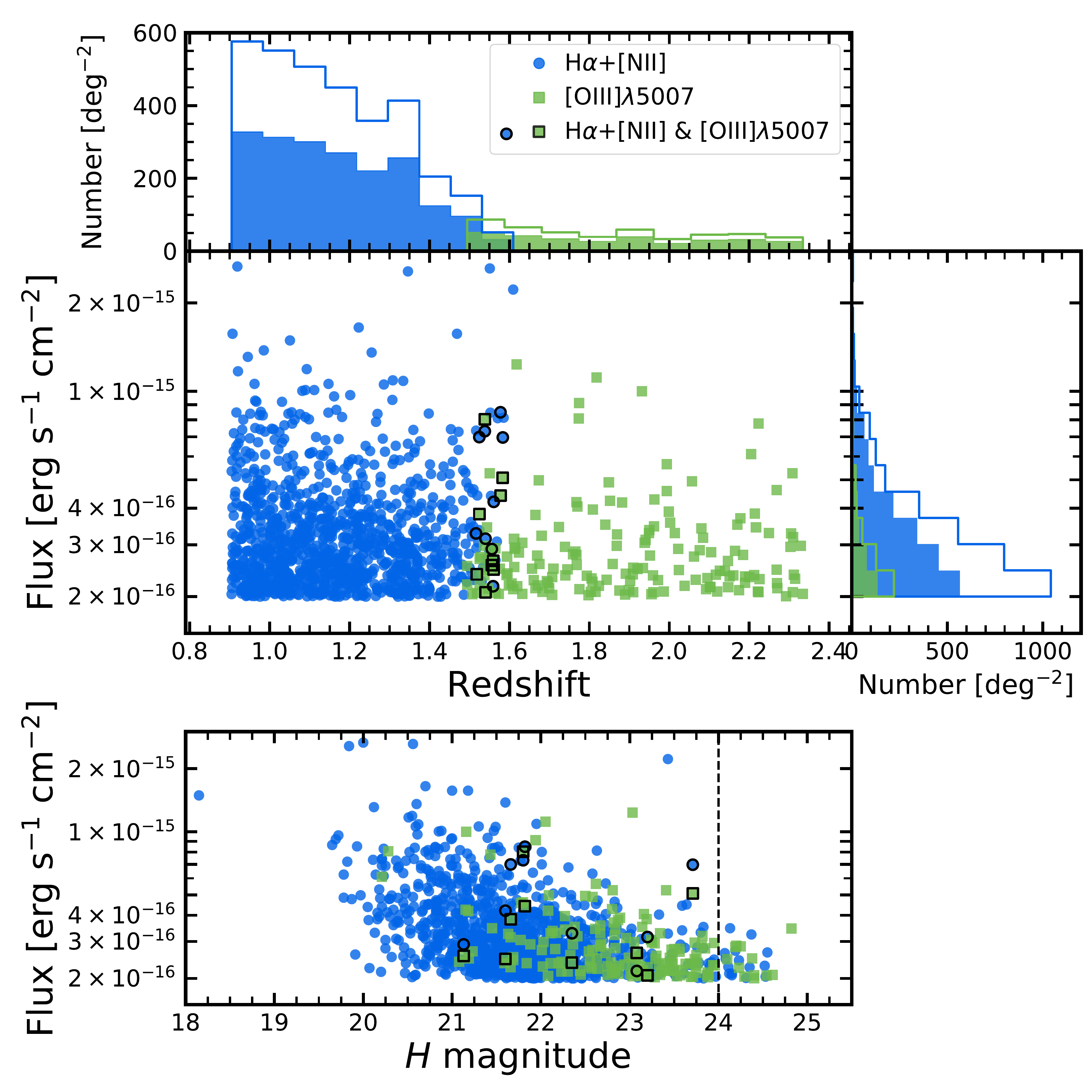}
\caption{
The \han\ and \oiiib\ line fluxes of sources in the Wide Sample (WS)
as a function of redshift (top panel) and $H$ magnitude (bottom panel). 
The handful of sources with both lines detected at the redshift and depth 
of the Euclid Wide Survey are outlined in black. The majority of 
sources in the WS are \han-emitters at $z\lesssim1.5$. 
The observed and completeness-corrected distributions of source redshift 
(top) and line flux (right) are shown as filled and empty histograms, 
respectively. 
In the bottom panel, the sources fainter than $H=24$ (black dashed line)
will be missed if Euclid spectral extraction is only performed 
for sources with $H < 24$. This subset amounts to $\sim$50 ELGs deg$^{-2}$,
or $\sim$2\% of the WS.
\label{fig:sample}}
\end{figure}

The WS consists of 1277 ELGs (2270 deg$^{-2}$), 
the majority of which are \han-emitters below redshift 
$z\lesssim 1.5$ (85\%, see Figure~\ref{fig:sample}). 
There are 73 galaxies in the redshift range $1.5 \lesssim z \lesssim 1.6$, 
where both \ha\ and \oiii\ are accessible to the Euclid Red grism. Of these, 
only 9 (16 deg$^{-2}$) have both \han\ and \oiiib\ bright enough for 
the Euclid WS selection. 
The median \han/\oiiib\ of these 9 galaxies is $1.45\pm0.30$, though the 
strength of this ratio increases with \han\ line flux, as can be seen 
in \cite{colbert2013} and \cite{mehta2015}.
The Euclid NISP instrument will reach a 5$\sigma$ sensitivity of 
24th magnitude in all three of its imaging filters. Fainter sources will 
be in the photometric catalogs, but the current observing strategy 
calls for spectral extraction only for sources brighter than this 
5$\sigma$ limit.
We note that the 28 real ELGs (50 deg$^{-2}$, 
$\sim$2\% of the WS) with $H > 24$ 
in the bottom panel of Figure~\ref{fig:sample}, all with emission 
lines brighter than the Euclid flux limit, would be missed by this extraction 
strategy. 
Extracting spectra for sources detected at lower S/N (e.g., $3-3.5\sigma$)
or down to fainter magnitudes ($H<24.5$) would allow for 
the recovery of these high-EW sources. Yet this strategy would also result 
in significantly more spectra to process and search for emission lines. 
For example, 10\% of sources in the full WISP+3D-HST catalog have 
continuum magnitudes in the range $24 < H <24.5$, amounting to 
$\sim$2500 more extracted spectra per square degree.

%%%%%%%%%%%%%%%%%%%%%%%%%%%%%%%%%%%%%%%%
\section{Emission Size Measurements} \label{sec:emissionsizes}
We aim to use \hst\ grism observations to predict the distribution of 
emission sizes that Euclid will detect as well as the effect source size will 
have on the Euclid selection function. 
Observations of a source, and therefore any resulting measurements of the 
source size and shape, are the result of the convolution of the intrinsic 
source shape and the point spread function (PSF) of the telescope and 
instrument. 
Before analyzing the size distributions, we must first deconvolve the 
observations with the PSF in order to recover the intrinsic sizes of the 
sources in the WISP and \tdhst\ catalogs.
In Section~\ref{sec:psf} we describe the construction of an 
empirical PSF for each imaging filter and as a function of wavelength for 
the grisms. 
We then present the methods for measuring the emission sizes in 
Section~\ref{sec:sizemodels}.

%%%%%%%%%%%%%%%%%%%%%%%%%%%%%%%%%%%%%%%%
\subsection{Constructing an empirical PSF} \label{sec:psf}
We construct an empirical PSF using the imaging and spectral stamps of 
$\sim$3000 stars in the WISP fields included in the WS. 
The stars are selected by $H$ band magnitude and half-light radius as 
described in Bruton et al. (\textit{in prep}).
We consider stars in the magnitude range $22 \geq H \geq 18.3$, where the 
faint limit is imposed to avoid selecting compact galaxies, and the upper limit 
conservatively removes stars that may be saturated or approaching the 
non-linearity regime of the detector where the \se\ centroids are unreliable. 
We do not explicitly select isolated sources, which are ideal for minimizing 
imaging and spectral overlap with nearby sources, but instead depend on the 
median profile to provide an accurate representation of the observed PSF.

\begin{figure}
\plotone{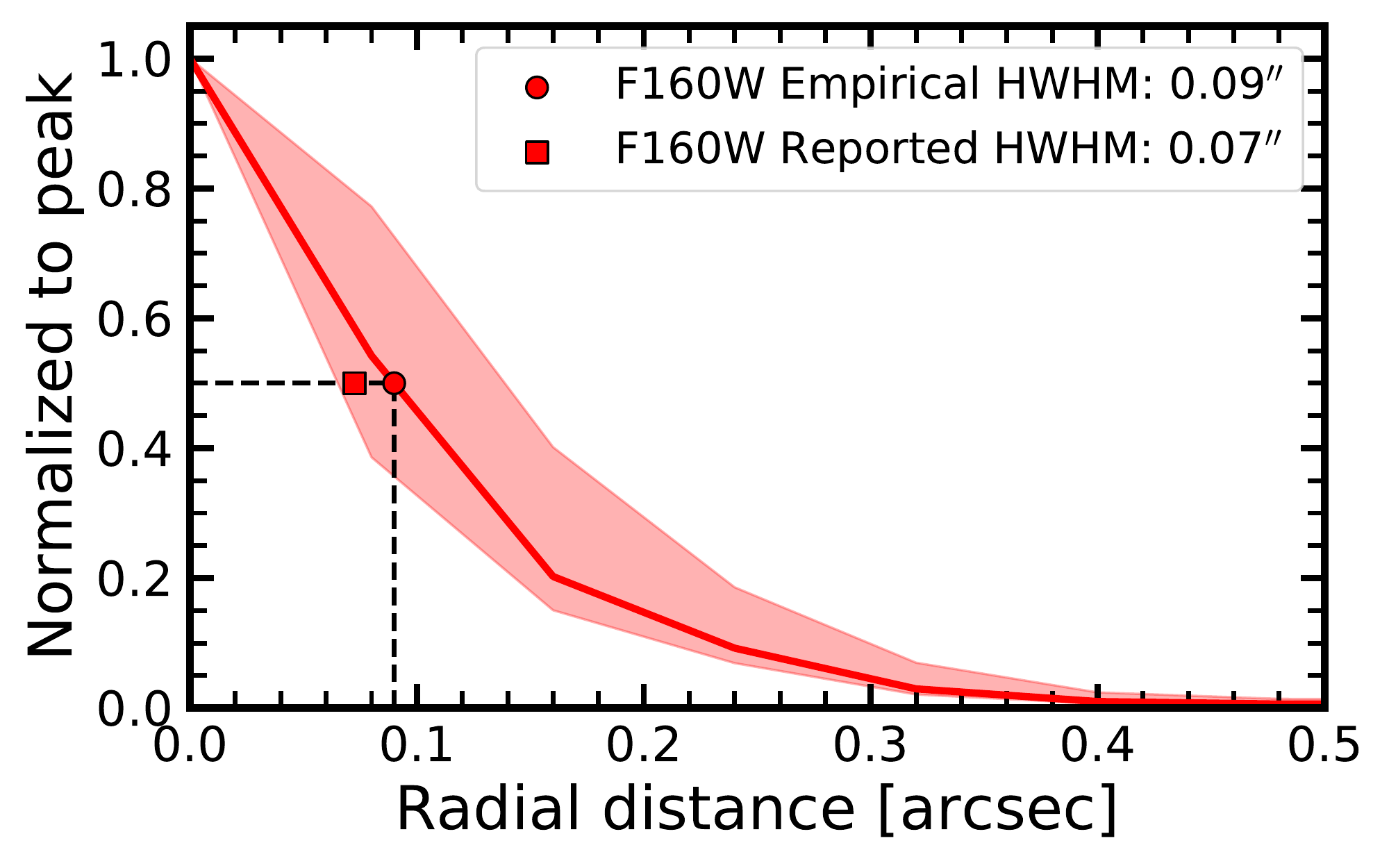}
\caption{The empirical \hst\ WFC3 F160W PSF measured using $\sim$1720 stars. 
The median azimuthally averaged radial profile is shown as the red curve, 
and the shaded band includes $\pm1\sigma$ of all measured profiles. 
The half width at half maximum is indicated by the circle and dashed lines. 
The measured HWHM is larger than that reported in the WFC3 Handbook (square). 
\label{fig:f160_psf}}
\end{figure}

We begin by describing the creation of the imaging PSF. For each star in the 
\se\ imaging catalog, we create $10\arcsec\times10\arcsec$ stamps in all 
available IR filters.
We then construct a radial profile of each star by calculating the 
azimuthally averaged flux in circular annuli of increasing radii. 
The median radial profile for F160W 
is shown in Figure~\ref{fig:f160_psf} as an example. The half width at 
half maximum (HWHM) is indicated by the circle and dashed lines and corresponds to 
a FWHM$=0\farcs18$, larger than the FWHM reported in the WFC3 Instrument 
Handbook\footnote{\url{www.stsci.edu/hst/wfc3/documents/handbooks/currentIHB}}
for Cycle 26: $0\farcs145$ for F160W. As we have measured the radial profile 
of each star individually, rather than from a stacked image, we conclude that 
the discrepancy is not caused by problems centering the stars in the imaging 
stamps.
There are not enough stars in all fields containing ELGs 
to measure a field-dependent PSF. Additionally, the \hst\ PSF is undersampled. 
The FWHM of an undersampled PSF is typically measured by sampling the PSF
with multiple stars and therefore multiple sub-pixel centroid positions. 
We therefore take the median profile as the effective PSF and adopt this  
FWHM for all ELGs, including those in \tdhst\ fields.
We note that the values reported in the handbook are listed before 
pixelation and are therefore expected to be smaller than the measurements 
of the pixelated PSF we perform here, though slight variations in the 
telescope focus during these observations can also contribute to the 
discrepancy. 
Here we aim to deconvolve the PSF from galaxy emission size measurements, 
and so adopt the larger, empirically--measured FWHMs to ensure the 
galaxy emission and PSF are measured consistently from the same data. 

\begin{figure*}
\plotone{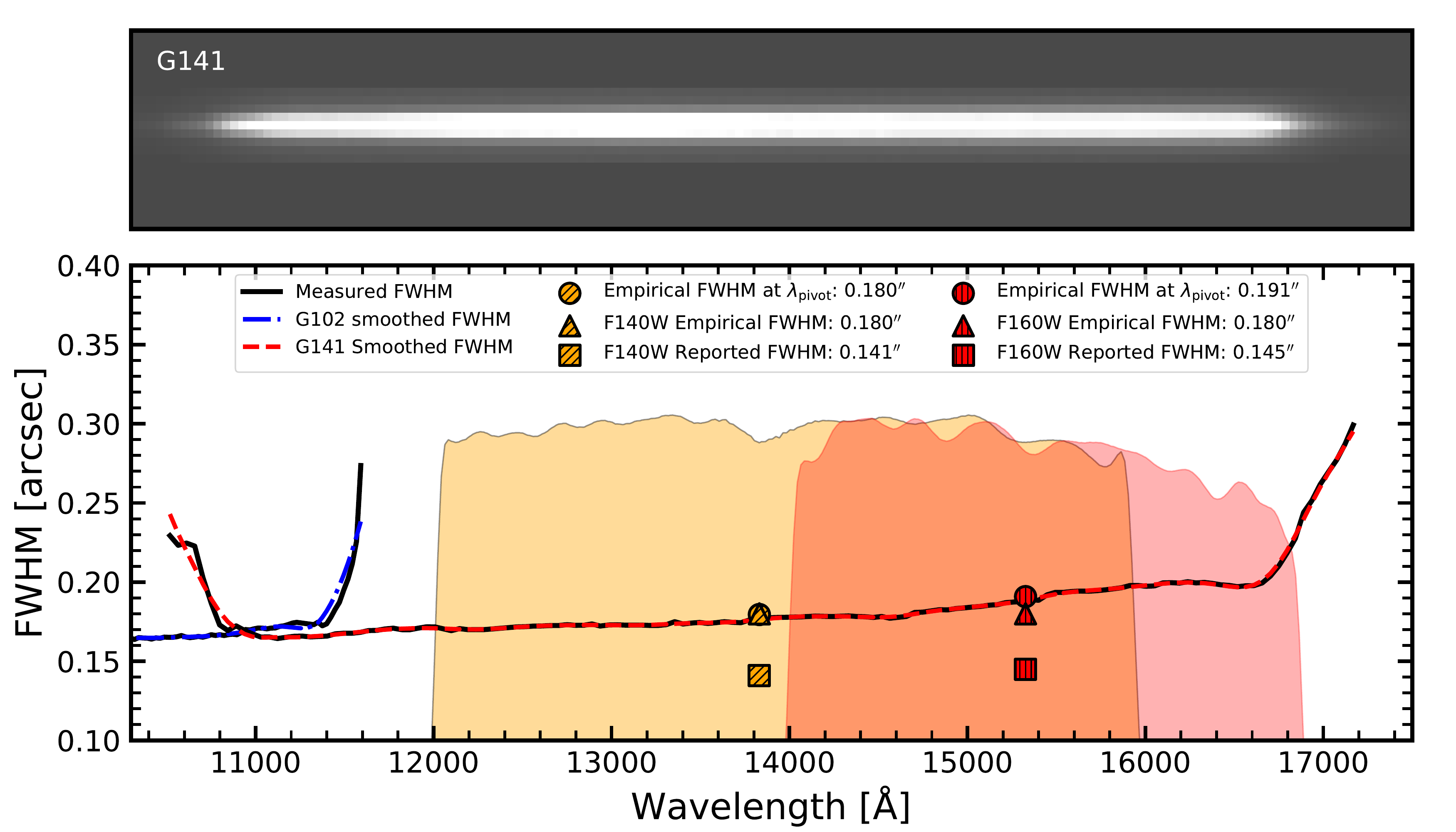}
\caption{The empirical G141 PSF, measured from the stacked spectra of 
$\sim$2700 stars. The stacked spectrum is shown in the top panel. 
The measured FWHM as a 
function of wavelength is shown in black for both grisms. 
The smoothed wavelength-dependent FWHM is shown in red (blue) for G141 (G102). 
The FWHMs integrated over the F140W and F160W filter profiles are plotted as 
orange and red circles, respectively, at the filter pivot wavelengths. 
These values are consistent with the FWHMs we measure for each filter in 
the imaging stamps (triangles), while the 
FWHM reported in the WFC3 Handbook for Cycle 26 (squares) are both lower by 
a factor of $\sim$0.2$-$0.3. We focus here on the filters used in the 
analysis of the Euclid WS: F140W, F160W, and G141. The PSF FWHMs for all 
filters, including F110W and G102, are listed in Table~\ref{tab:psfs}. 
\label{fig:grism_psf}}
\end{figure*}

The grism PSFs are measured on median-combined spectral stamps in order to 
achieve a high S/N. 
As \texttt{aXe} drizzles together the individual exposures using the positions 
of the sources in the corresponding individual imaging exposures, the spatial 
centroid and the wavelength solution are consistent enough in each spectral 
stamp to allow stacking. 
The combined stellar spectrum in G141 is displayed in the top panel of 
Figure~\ref{fig:grism_psf}. We measure the FWHM of the combined spectrum along 
the spatial axis (vertically in Figure~\ref{fig:grism_psf}) by fitting a 
Gaussian to the flux profile at each wavelength in a moving average window 
5 pixels wide. The FWHM measured in this manner is plotted as a function of 
wavelength in the bottom panel of Figure~\ref{fig:grism_psf} (black curve). 
We smooth the wavelength-dependent FWHM using a Savitzky-Golay filter 
\citep{savgol1964} with a window 11 pixels wide (red curve). 
Finally, we calculate the integrated FWHM over the passband of each imaging
filter and confirm that the FWHMs measured in the grisms are consistent with 
those measured in imaging. This comparison is also displayed in 
Figure~\ref{fig:grism_psf} for G141, F140W, and F160W. 
Table~\ref{tab:psfs} provides the measured FWHM in each filter as well as 
the number of stars that were included in the measurement.

\begin{deluxetable}{c|cccc}
\centering
\tablewidth{0pt}
\tablecaption{Empirical WFC3 PSFs \label{tab:psfs}}
\tablehead{
\colhead{Filter} & \colhead{$N_{\mathrm{stars}}$} & 
\colhead{Measured FWHM} & \colhead{Reported FWHM} \\
\colhead{} & \colhead{} & 
\colhead{[arcsec]} & \colhead{[arcsec]} 
}
\startdata
F110W & 1408 & 0.207 & 0.130 \\ 
F140W & 916  & 0.180 & 0.141 \\ 
F160W & 1720  & 0.180 & 0.145 \\
G102 & 1523  & 0.164 & 0.128 \\
G141 & 2749  & 0.178 & 0.141 \\
\enddata
\tablecomments{The reported FWHM are taken from the WFC3 Handbook
and represent the measurement of the PSF pre-pixelation and at the
wavelengths that most closely match the pivot wavelengths of the filters.
The grism FWHMs are those for the approximate midpoint 
wavelengths: 10000~\AA\ for G102 and 14000~\AA\ for G141.}
\end{deluxetable}

%%%%%%%%%%%%%%%%%%%%%%%%%%%%%%%%%%%%%%%%
\subsection{Modeling Continuum and Line Emission} \label{sec:sizemodels}
\begin{figure}
\epsscale{1.1}
\plotone{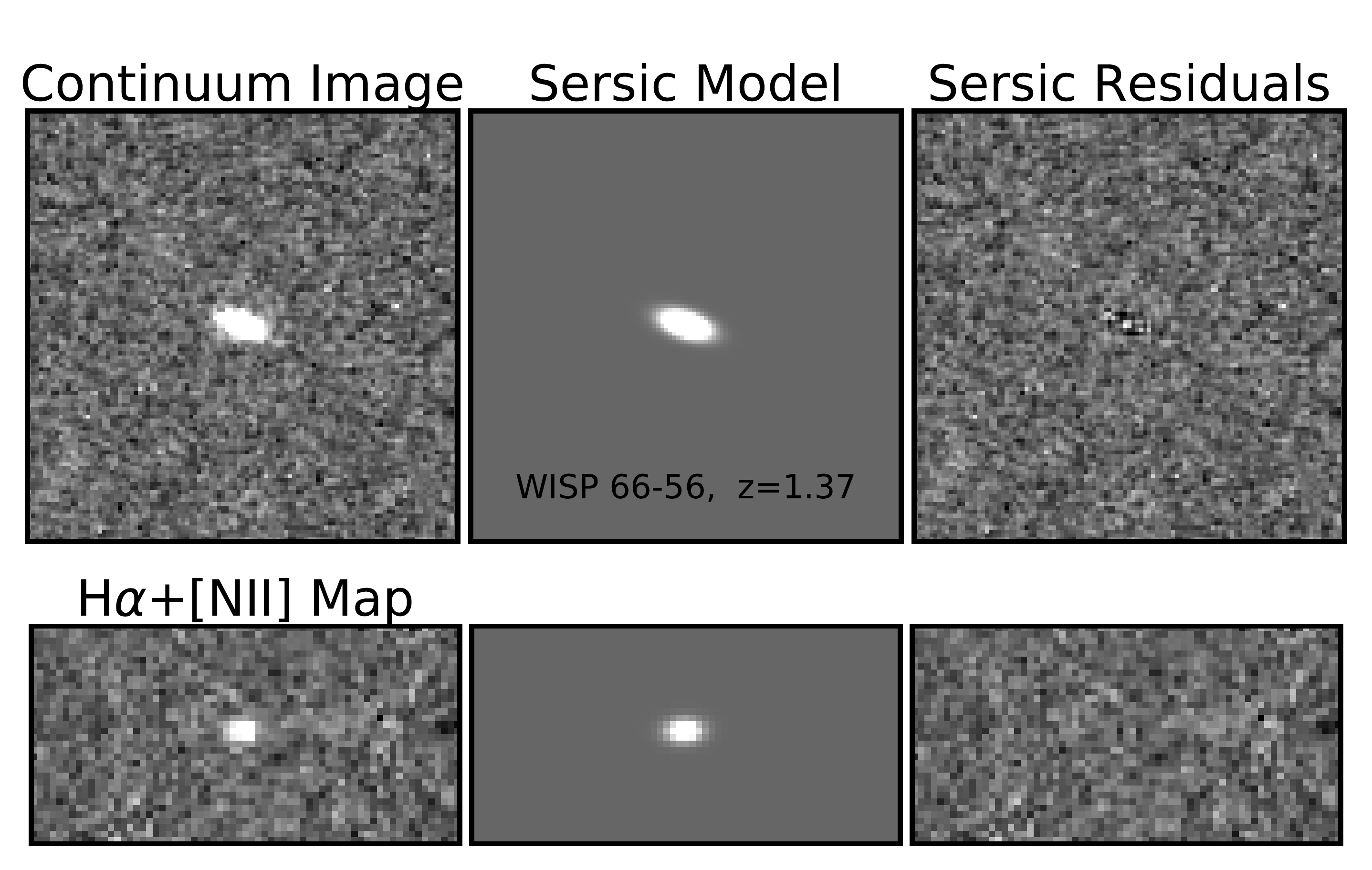}
\caption{
The best-fit S\'{e}rsic models for the continuum image (top row) 
and \ha\ emission line map (bottom row)
for an example source. The columns from left to right show the input image 
stamps, the S\'{e}rsic model, and the residuals. 
\label{fig:emissionmodel}}
\end{figure}

We measure the sizes of each ELG in both the continuum and the \han\ emission.
The continuum sizes are measured on $9^{\prime\prime} \times 9^{\prime\prime}$ 
stamps created from the $H$ band direct images in either the F140W or F160W 
filters. The emission line sizes are measured on stamps created from the
two-dimensional spectra extracted from the full grism images 
(see Section~\ref{sec:wisp} for a description of this spectral extraction).
We create stamps for each emission line from the 2D spectra as follows. 
The stamps extend 35 pixels in the wavelength direction ($\sim$850~\AA\ 
in G102, $\sim$1600~\AA\ in G141) on either side of the center of the 
emission line. We fit the continuum row by row in the stamp by fitting 
a line to the fluxes in the pixels on either side of the line excluding 
8 pixels (370~\AA) centered at the wavelength of the emission line.
We subtract each linear
fit from the corresponding full row and are left with a 
continuum-subtracted map of each galaxy in the given emission line. 
An example of an \han\ emission line map is shown in 
the bottom left panel of Figure~\ref{fig:emissionmodel}.

Next, we model the shapes of the continuum and \han\ emission for the 
Euclid WS sources using S\'{e}rsic profiles. 
The S\'{e}rsic profile describes the intensity of the source as a function 
of radius \citep{sersic1963,sersic1968}. The functional form is given by:
\begin{equation}
I(R) = I_{e} \ \mathrm{exp} \left\{ -b_n \left[ \left(\frac{r}{r_{e}} \right) ^{1/n} - 1 \right] \right\},
\end{equation}
where $r_e$ is an effective or scale radius and $I_e$ is the profile 
intensity at $r_e$. The S\'{e}rsic index, $n$, determines the 
shape of the light profile, with larger values corresponding to more 
centrally concentrated sources. A value of $n=1$ results in an exponential 
profile that is a good approximation of disk galaxies, while $n=4$ 
gives the \cite{vaucouleurs1948} profile approximating elliptical galaxies. 
The constant $b_n$ is coupled to $n$ such that $r_e$ is the half-light 
radius --- or the radius that encircles half of the 
light emitted by the source --- and is therefore not a free parameter.

For each image stamp and emission line map, we determine the best-fit 
S\'{e}rsic models using the two-dimensional image fitting software 
\galfit\footnote{\url{https://users.obs.carnegiescience.edu/peng/work/galfit/galfit.html}} \citep[v3.0; ][]{peng2010}.
In fitting, \galfit\ convolves the S\'{e}rsic profiles with a Gaussian 
kernel to emulate the PSF, such that the best-fit model parameters will be 
those of the PSF-corrected emission shapes and sizes.
For the continuum emission measured in the imaging stamps, 
the FWHM of either the F140W or F160W filter is used. 
The FWHM for each emission line map is taken from the smoothed function 
described in Section~\ref{sec:psf} at the wavelength of the line. 
We use relatively large stamp sizes ($9\arcsec$ in the continuum) so 
that a sufficient number of sky pixels are available for the \galfit\ fitting 
algorithm. However, we ensure that close neighboring sources do not interfere 
with the fitting of the target source by constraining all models to have 
centroids within $\pm3$ pixels of the stamp centers.
The stamps, models, and residuals for one of the WISP sources are shown in 
Figure~\ref{fig:emissionmodel} as an example of the model fitting. 

We perform the same size measurement on the simulated data discussed in 
Section~\ref{sec:completeness} and Appendix~\ref{app:wispcorrections}. Recall that the simulated sources are the 
same size and shape in both the continuum and emission lines. The 
effective radii should therefore be tightly correlated, and we can use 
the scatter as an estimate of the statistical error of our model fitting. 
As the synthetic sources were simulated as two-dimensional Gaussians,  
we similarly fit the simulated data with elliptical Gaussian models rather 
than the S\'{e}rsic profiles used for the real sources.

\begin{figure}
\epsscale{1.1}
\plotone{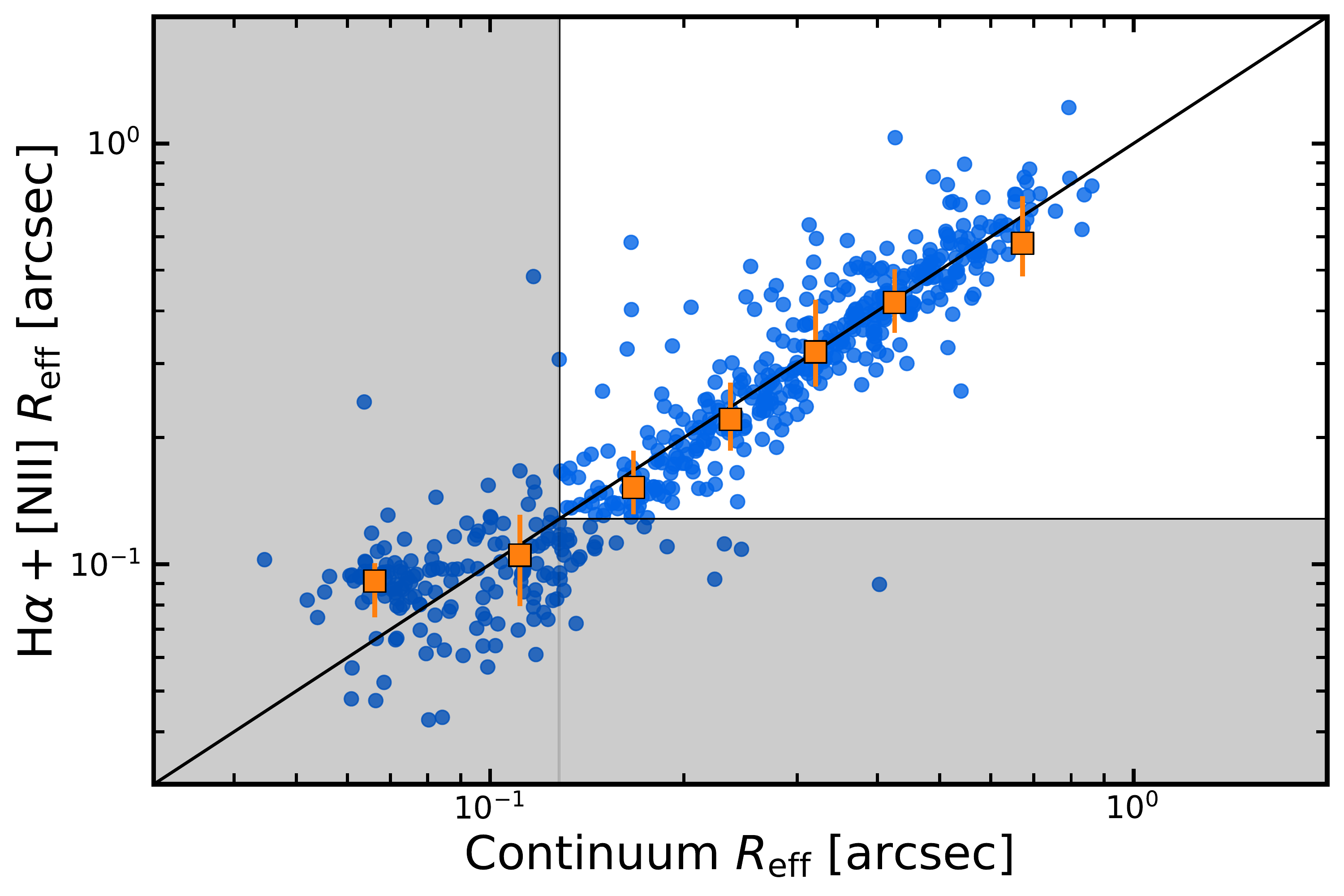}
\caption{
The effective radii of simulated sources measured in both the continuum 
and \han\ emission. The orange squares show the median values and $1\sigma$ 
scatter in bins of equal number of sources.
Simulated sources are both created by and fit with 
elliptical Gaussian models. Since \texttt{aXeSIM} creates sources that are 
the same size in both the continuum and line emission, we take the 
$\sim0\farcs05-0\farcs15$ scatter in this relationship as an estimate of the 
statistical error of our model fitting.
The shaded area indicates the size of one pixel in the WISP images, 
where the pixel scale is $0\farcs13$/pixel. 
\label{fig:simsizes}}
\end{figure}

The \reff\ for the simulated data are shown in Figure~\ref{fig:simsizes}, 
where here \reff\ refers to a circularized radius constructed from the 
standard deviation of the Gaussian model 
along each axis, \reff~$=\sqrt{\sigma_x \sigma_y}$.
The median \han\ \reff\ in bins of continuum \reff\ are plotted as squares 
with $1\sigma$ error bars. 
The standard deviation of the relation between the continuum and emission 
line $R_{\mathrm{eff}}$ is $\sim0\farcs05-0\farcs15$. 
The continuum and \han\ emission sizes are correlated down to small radii, 
\reff$\sim0\farcs07$, below which the \reff\ are smaller than one pixel in 
the grism spectra and therefore unreliable. 
We present the relationship between continuum and \han\ \reff\ for the 
observed WS sources in Section~\ref{sec:hasize}.

%%%%%%%%%%%%%%%%%%%%%%%%%%%%%%%%%%%%%%%%
\section{Results and Discussion} \label{sec:results}

%%%%%%%%%%%%%%%%%%%%%%%%%%%%%%%%%%%%%%%%
\subsection{Emission Line Number Counts} \label{sec:counts}
We begin by considering the number of ELGs that meet the selection criteria
for the Euclid WS. 
Galaxies emitting \han\ are the main target for the dark energy science, 
as they will be used to trace the large scale structure at $z\sim1-2$.
There are $1939\pm21$ \han-emitters deg$^{-2}$ in the
WS from $0.9 \leq z \leq 1.6$ and an additional $288\pm9$ \oiiib-emitters
deg$^{-2}$ up to $z \sim 2.3$. 
Correcting these observed counts for the incompleteness of the WFC3 grism 
data, there are 3266 (\han) and 445 (\oiiib) deg$^{-2}$, respectively.
In addition to the WS sources with \han\ and \oiiib\ emission, there are 
a handful of sources at lower redshift ($z\sim$0.4) that were selected due 
to the strength of the \siii\ and \hei\ emission. The number counts with and 
without completeness corrections of all selected emission lines are 
presented in Table~\ref{tab:counts}. 

\begingroup
\renewcommand*{\arraystretch}{1.3}
\begin{deluxetable}{c|cccc}
\centering
\tablewidth{0pt}
\tablecaption{WS number counts for lines with $1.25 \leq \lambda_{\mathrm{obs}} \lesssim 1.7$ \micron\ \label{tab:counts}}
\tabletypesize{\footnotesize}
\tablehead{
\colhead{} & \colhead{Flux} & \colhead{N$_{\mathrm{obs}}$} & 
\colhead{N$_{\mathrm{obs}}$/deg$^2$} & \colhead{N$_{\mathrm{corr}}$/deg$^2$} 
}
\startdata
%\arraystretch{3}
\han\ & $\geq4$ & $269^{+6.2}_{-7.0}$  & $478.1^{+10.9}_{-12.4}$  & $704.5^{+19.7}_{-18.0}$  \\
{\footnotesize ($0.9 \leq z \lesssim 1.6$)} & $\geq3$ & $516\pm 10.0$  & $917.2\pm17.8$  & $1421.0^{+42.0}_{-35.0}$  \\
& $\geq2$ & $1091\pm 12.0$  & $1939.2\pm21.3$  & $3266.0^{+157.7}_{-174.8}$  \\
& $\geq1$ & $2378^{+17.8}_{-19.7}$  & $4226.7^{+31.7}_{-34.9}$  & $7887.3^{+148.5}_{-166.0}$  \\[2pt]
\hline 
\oiiib\ & $\geq4$ & $20^{+1.0}_{-2.0}$  & $35.5^{+1.8}_{-3.6}$  & $46.0^{+2.2}_{-2.0}$  \\
{\footnotesize ($1.5 \leq z \lesssim 2.4$)} & $\geq3$ & $46^{+2.0}_{-2.2}$  & $81.8^{+3.6}_{-3.8}$  & $112.2^{+3.8}_{-2.5}$  \\
& $\geq2$ & $162\pm 5.0$  & $287.9\pm8.9$  & $444.6^{+15.5}_{-10.6}$  \\
& $\geq1$ & $517^{+8.3}_{-9.0}$  & $918.9^{+14.8}_{-16.0}$  &$1608.5^{+14.6}_{-11.9}$  \\[2pt]
\hline
\siii~$\lambda9069$ & $\geq4$ & $6\pm1.0$ & $10.7\pm1.8$  & $14.6\pm1.3$  \\
{\footnotesize ($0.38 \leq z \lesssim 0.87$)} & $\geq3$ & $6\pm1.0$  & $10.7\pm1.8$  & $14.6\pm1.3$  \\
& $\geq2$ & $6^{+1.0}_{-2.0}$  & $10.7^{+1.8}_{-3.6}$  & $14.6\pm1.3$  \\
& $\geq1$ & $6\pm2.0$  & $10.7\pm3.6$  & $14.6\pm1.3$  \\[2pt]
\hline
\siii~$\lambda9532$ & $\geq4$ & $4\pm1.0$  & $7.1\pm1.8$  & $11.2^{+1.6}_{-1.2}$  \\
{\footnotesize ($0.31 \leq z \lesssim 0.78$)} & $\geq3$ & $6^{+1.0}_{-2.0}$  & $10.7^{+1.8}_{-3.6}$  & $15.7^{+1.5}_{-1.4}$  \\
& $\geq2$ & $10^{+3.0}_{-2.0}$  & $17.8^{+5.3}_{-3.6}$  & $25.1^{+1.9}_{-1.3}$  \\
& $\geq1$ & $20\pm3.0$  & $35.5\pm5.3$  & $52.6^{+1.7}_{-1.5}$  \\[2pt]
\hline
\hei~$\lambda10830$ & $\geq4$ & $3\pm1.0$  & $5.3\pm1.8$  & $7.9^{+1.8}_{-1.4}$  \\
{\footnotesize ($0.15 \leq z \lesssim 0.57$)} & $\geq3$ & $6^{+2.0}_{-1.0}$  & $10.7^{+3.6}_{-1.8}$  & $18.8^{+3.9}_{-2.3}$  \\
& $\geq2$ & $8^{+2.0}_{-1.2}$  & $14.2^{+3.6}_{-2.1}$  & $28.9^{+7.2}_{-4.0}$  \\
& $\geq1$ & $10\pm2.0$  & $17.8\pm3.6$  & $35.6^{+5.1}_{-6.4}$  \\
\enddata
\tablecomments{Observed (N$_{\mathrm{obs}}$) and completeness-corrected 
(N$_{\mathrm{corr}}$) cumulative number counts.
Numbers presented here are for unique sources, i.e., ELGs 
with both \han\ and \oiiib\ are counted only for \han. The redshift ranges
associated with each emission line are given for observed wavelengths 
$12500 \leq \lambda_{\mathrm{obs}} \lesssim 17000$~\AA, where the approximate 
upper limit is set by the decreasing sensitivity of the \hst\ G141 grism.
The line fluxes indicated in the first column are $\times 10^{-16}$ \esc.}
\end{deluxetable}
\endgroup

We obtained the errors presented on the number counts in 
Table~\ref{tab:counts} through a Monte Carlo process by creating 200 
realizations of the full WISP$+$3D-HST emission line catalog, re-running the 
WS sample selection, and measuring the resulting distribution of number 
counts. For the observed number counts 
($N_{\mathrm{obs}}$), each catalog realization is generated with redshifts
and emission line fluxes (and therefore EWs) pulled randomly from Gaussian 
distributions centered at the measured values and with standard deviations 
equal to the uncertainties on these measurements in the catalog. 
The number of sources recovered by the selection criteria varies from 
realization to realization.
For the completeness-corrected number counts ($N_{\mathrm{corr}}$), we leave 
the line fluxes and redshifts untouched and pull the completeness corrections 
from Gaussian distributions with standard deviations equal to the uncertainties
on the completeness corrections. In this case, the number of recovered 
sources stays the same, while the completeness-corrected number varies.
In both cases, we report the \nth{16} and \nth{84} percentiles as the lower 
and upper errors in Table~\ref{tab:counts}, respectively.
\begin{figure*}
\epsscale{1.1}
\plotone{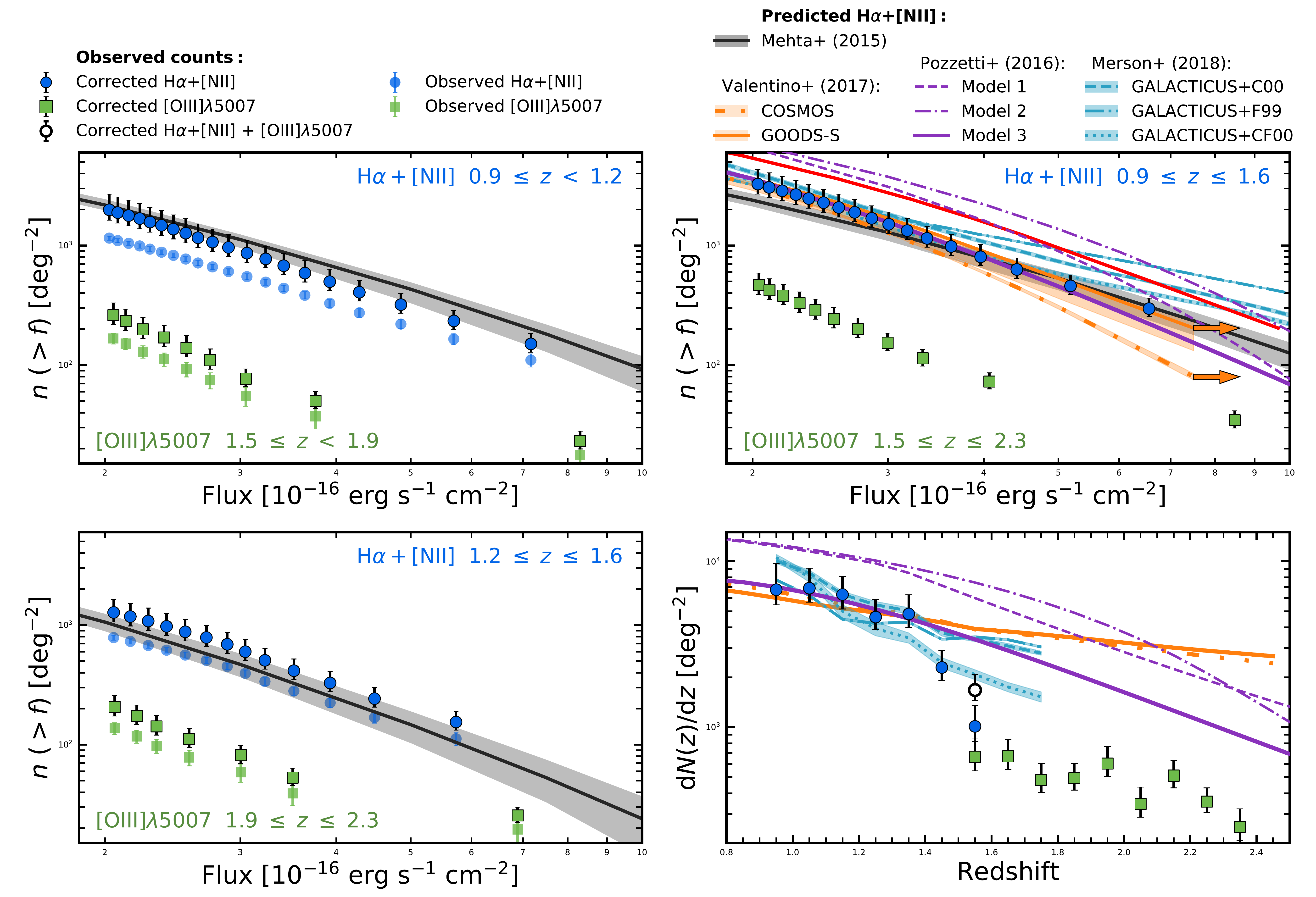}
\caption{
The cumulative number counts (left column and top right panel) 
and redshift distribution (bottom right)
of \han\ and \oiiib-emitters in the WS.
In the left column we have separated the cumulative number counts into two 
redshift 
bins to demonstrate the evolution in the number counts of each line with 
redshift. The number counts of \han\ and \oiiib-emitters are shown as 
blue circles and green squares, respectively. 
The fainter points are the observed number counts and the larger, solid 
symbols indicate those that have been corrected for survey incompleteness. 
The black curve indicates the \han\ predictions from \cite{mehta2015} 
in each redshift bin.
In the top right panel, we compare the observed, completeness-corrected
\han\ number counts across the full redshift range to those predicted by 
\cite{mehta2015} (black curve); the models from \cite{pozzetti2016} 
(purple curves); \cite{valentino2017} for galaxies in the COSMOS 
(dot-dashed orange) and GOODS-S (solid orange) fields;
and \cite{merson2018} for dust models
from \cite{calzetti2000} (C00; dashed blue),
\cite{ferrara1999} (F99; dot-dashed blue), 
and \cite{charlot2000} (CF00; dotted blue). 
The arrows indicate the level of uncertainty in number counts 
associated with the choice of \nii\ correction (see the text for details).
The redshift distributions ($dN/dz$) of the WS ELGs are shown in the bottom 
right panel, again compared with the models from \cite{pozzetti2016},
\cite{valentino2017}, and 
\cite{merson2018}. In the redshift bin where both \han\ and \oiiib\ are 
both accessible to the WFC3 grism, the total (\han)+(\oiiib) counts are 
indicated by an empty circle.
\label{fig:ndensity}}
\end{figure*}

The cumulative number counts of both \han-emitters and
\oiiib-emitters are shown in Figure~\ref{fig:ndensity}. 
The observed counts are shown as fainter points with Poisson 
uncertainties determined by the number of sources in each bin. The 
completeness-corrected counts are calculated as 
\begin{equation}
N_{\mathrm{corr}} = \displaystyle\sum_i \frac{1}{C_i},
\end{equation}
where $C_i$ is the completeness for each source in the bin. The error bars
are obtained by varying the completeness corrections maximally within the 
uncertainties, i.e. $C_i+\sigma_{C_i}$ and $C_i - \sigma_{C_i}$. These error 
bars therefore represent the range of possible number densities given the 
uncertainties on the completeness corrections.
The number counts are separated into two redshift bins in the left column 
to highlight the evolution in the
number density with redshift due to the increasing luminosity limit.
There is a factor of more than 1.5 times more \han-emitters at $0.9\leq z<1.2$ 
than at $1.2 \leq z \leq 1.6$.
The observed number counts are in good agreement with the predictions 
from \cite{mehta2015}, calculated using a subset of WISP fields from the 
earlier data reduction and original line finding procedure. 
The top right panel includes the cumulative number counts for the full 
redshift ranges available to the \hst\ grism: 
$0.9 \leq z_{\mathrm{H}\alpha \mathrm{+[NII]}} \lesssim 1.6$ and 
$1.9 \leq z_{\mathrm{[OIII]}\lambda 5007} \lesssim 2.3$. Finally, the 
redshift distributions of \han\ and \oiiib-emitting galaxies 
($\mathrm{d}N/\mathrm{d}z$) in bins of $\Delta z=0.1$ are shown in the 
bottom right panel.

In the right column of Figure~\ref{fig:ndensity}, we compare the observed 
\han\ number counts with the empirical models of \citet{pozzetti2016} in 
purple. The three models represent different parameterizations of the \ha\ 
luminosity function and its redshift evolution. 
For the purposes of comparison, we have converted the \ha\ counts of 
all three models to \han\ counts using a 
fixed \nii/\ha\ line ratio: $\mathrm{H}\alpha=0.71$~(\han), 
the same conversion used in Section~5 of \citet{pozzetti2016} while comparing 
the model counts to observations.
The observed \han\ number counts in the top right panel of 
Figure~\ref{fig:ndensity} agree most closely with Model 3, which is the 
result of a fit to observations presented by 
\citet[][HiZELS]{sobral2013}, \citet[][WISP]{colbert2013}, \cite{yan1999}, 
and \cite{shim2009}. Model 3 is also one of the models against which the 
Euclid Flagship mock catalog has been calibrated. While the cumulative 
counts agree with Model 3 at almost all fluxes across the full redshift 
range, the distribution of \han-emitters with redshift falls off 
at $z\sim1.5$. For \han, this redshift corresponds approximately to the 
wavelengths at which the sensitivity in the G141 grism begins to decrease. 

We also show the predictions from \cite{valentino2017} in orange in 
the right panels of Figure~\ref{fig:ndensity}. 
\cite{valentino2017} use the large photometric samples in the COSMOS and 
GOODS-S fields to predict the number counts of ELGs that will be accessible 
to future galaxy redshift surveys. They derive \ha\ fluxes from the 
star formation rates obtained via spectral energy distribution fitting
and use \ha-emitters at $z=1.55$ observed with 
the FMOS-COSMOS survey \citep{silverman2015} to calibrate the star 
formation rate to \ha\ conversion. 
The \han\ predictions for galaxies on the star-forming main sequence in the 
range $0.9 \leq z \leq1.6$ are shown in orange in Figure~\ref{fig:ndensity}.
The counts have been corrected for the Eddington bias, which is a bias 
introduced by measurement uncertainties that can enhance the observed number of 
bright galaxies compared to fainter galaxies \citep{eddington1913}.
The orange shaded bands indicate the 68\% Poissonian confidence intervals, 
where the authors report the maximum of the upper and lower Poisson 
uncertainties. The uncertainties due to the Eddington bias correction are 
not included here, but are available in Table~3 of \cite{valentino2017}.
The cumulative flux counts, particularly in the GOODS-S field, 
are consistent with the \hst\ grism measurements 
at all fluxes, and are in good agreement with the predictions of 
\cite{mehta2015} and \cite{pozzetti2016}.
The redshift distributions of galaxies in both fields are consistent up to 
$z\sim1.5$, where the differential counts from \cite{valentino2017} 
follow a shallower evolution than those from the other predictive works. 

We note that the different \nii\ corrections adopted by each team 
can introduce some systematic uncertainties and contribute to this 
disagreement in the blended counts.
Specifically, while we have used a single correction for all number counts 
from \citet{pozzetti2016}, \citet{valentino2017} employ a complex correction 
as a function of mass that is smaller on average than that used for the
models from \citet{pozzetti2016}. As a result, the \ha-only counts (i.e. not 
including the \nii\ correction) from \cite{valentino2017} lie between Models 
1 and 3 of \citet{pozzetti2016}, yet their \han\ counts are generally lower 
and agree most closely with Model 3 (see Fig. 12 from \citealt{valentino2017} 
compared with the upper right panel of Figure~\ref{fig:ndensity}).
The median log$_{10}$(\nii/\ha) of the sample in \citet{valentino2017} 
is $\sim$$-0.45$ for galaxies with an \ha\ flux of $2\times10^{-16}$ \esc\
(see their Fig.~9), while the correction applied here to Models 1, 2, and 3 
corresponds to log$_{10}$(\nii/\ha$)=-0.39$. 
The arrows in the upper right panel of Figure~\ref{fig:ndensity} 
indicate the extent to which the number counts of 
\citet{valentino2017} would change if the \han\ fluxes were boosted by 
an additional factor (corresponding to $\Delta$log$_{10}$(\nii/\ha$)=0.06$)
to match the \nii/\ha\ ratio adopted by \citet{pozzetti2016}.
These arrows can be interpreted as the approximate uncertainty in number 
counts due to the \nii\ correction. As the \nii/\ha\ line ratio remains 
uncertain at these redshifts and observations at the resolution of the 
\hst\ grism cannot provide adequate constraints, these uncertainties are 
an important consideration when comparing \ha\ and \han\ number counts 
from different models and observations. 

We finally compare the observations with predictions from 
\cite{merson2018}, who use the 
\textsc{Galacticus} galaxy formation model \citep{benson2012} 
and the dust attenuation methods 
from \cite{ferrara1999}, \cite{calzetti2000}, and \cite{charlot2000} 
to predict \han\ number counts in the redshift range
$0.9\leq z \leq 1.55$, matching that available to the \hst\ grism. 
\cite{merson2018} use \han\ blended fluxes where the \nii/\ha\ ratios are 
determined by cross-matching the stellar mass and specific star formation 
rate of each \textsc{Galacticus} galaxy to the SDSS sample from
\cite{masters2016}. 
For each dust model, the red curve and shaded region in 
Figure~\ref{fig:ndensity} is the mean and standard deviation of 1000 
Monte Carlo realizations sampling the model's optical depth parameters.
The likelihoods for the sampling were constructed as 
$\mathcal{L} \propto \mathrm{exp}(-\chi^2/2)$, where the $\chi^2$ values were 
obtained by stepping through the dust parameter space and comparing 
\textsc{Galacticus} counts to the WISP counts from \cite{mehta2015} for 
$0.7 \leq z \leq 1.5$ \citep[see][for more information]{merson2018}.
While the \hst\ grism \han\ number counts presented in this paper (blue 
circles in Figure~\ref{fig:ndensity}) are lower than all three predictions 
from \cite{merson2018} for the brightest galaxies, the predictions and 
observations are consistent at the depth of the Euclid Wide Survey.

The number of galaxies observed by WISP$+$3D-HST are a lower limit to those 
that will be observed by Euclid. 
We remind the reader that the upper wavelength of the Euclid Red grism 
is $\sim$18500~\AA, 1500~\AA\ redder than the WFC3 G141 grism.
The Euclid \han\ and \oiiib\ observations will therefore extend out to 
$z\sim1.8$ and $z\sim2.7$, respectively. Euclid will detect more 
sources per square degree than those reported here.
However, as can be seen 
in Figure~\ref{fig:sample}, at the depth of the Euclid Wide Survey,
the number of detected ELGs drops quickly with redshift. 
The majority of the ELGs that Euclid 
detects will be \han\ at $z\lesssim1.5$, a population that is fully 
sampled by the \hst\ grism observations presented here.

With these \hst\ grism observations, we show that Euclid will meet the goal of measuring redshifts for $\sim$25 million galaxy redshifts 
over 15000 deg$^2$.
Extrapolating the completeness-corrected number densities to the full 
Euclid Wide Survey area provides a rough estimate of $\sim$48 million 
\han-emitters and $\sim$6 million \oiiib-emitters down to 
$f\geq 2\times 10^{-16}$ \esc. 
Even the observed, uncorrected counts, which provide an estimate in the case 
the Euclid galaxy redshift survey and these \hst\ grism observations 
suffer from the same level of incompleteness\footnote{However, with multiple 
roll angles planned for the Euclid grism observations, the incompleteness 
due to spectral confusion and emission lines lost to nearby bright neighbors 
will be lower than it is for the \hst\ grism data, particularly that of the 
WISP parallel data.}, are larger than the planned number of galaxy redshifts. 
These \hst\ observations therefore contribute a valuable resource  
to the effort to calibrate and verify the performance of 
the planned Euclid survey. 
Finally, we note that at the 
resolution of the Red grism, Euclid will be able to resolve the \ha\ and 
\nii~$\lambda\lambda6548,6584$ doublet for compact galaxies. However, the 
contribution from \nii\ for compact, low-mass ($<10^{10}$ $M_{\odot}$) 
galaxies is $\lesssim$10\% \citep[e.g.,][]{erb2006,masters2016,faisst2018}. 
Therefore, even for the galaxies for which Euclid will resolve the two lines,
we do not expect a \nii\ correction to the \han\ fluxes to 
significantly affect our results.

%%%%%%%%%%%%%%%%%%%%%%%%%%%%%%%%%%%%%%%%
\subsection{Emission and Continuum Sizes} \label{sec:sizes}
The location and size of the window used for spectral extraction from grism
images depend on the detection of the galaxy in a direct image.
This process relies on a few assumptions, including (1) that the full 
extent of the source has been detected in the direct image, and (2) that 
the emission line size is correlated with the source size in the direct image. 
We briefly explore both assumptions below.

%%%%%%%%%%%%%%%%%%%%%%%%%%%%%%%%%%%%%%%%
\subsubsection{Flux Loss from Spectral Extraction}
\begin{figure}
\epsscale{1.2}
\plotone{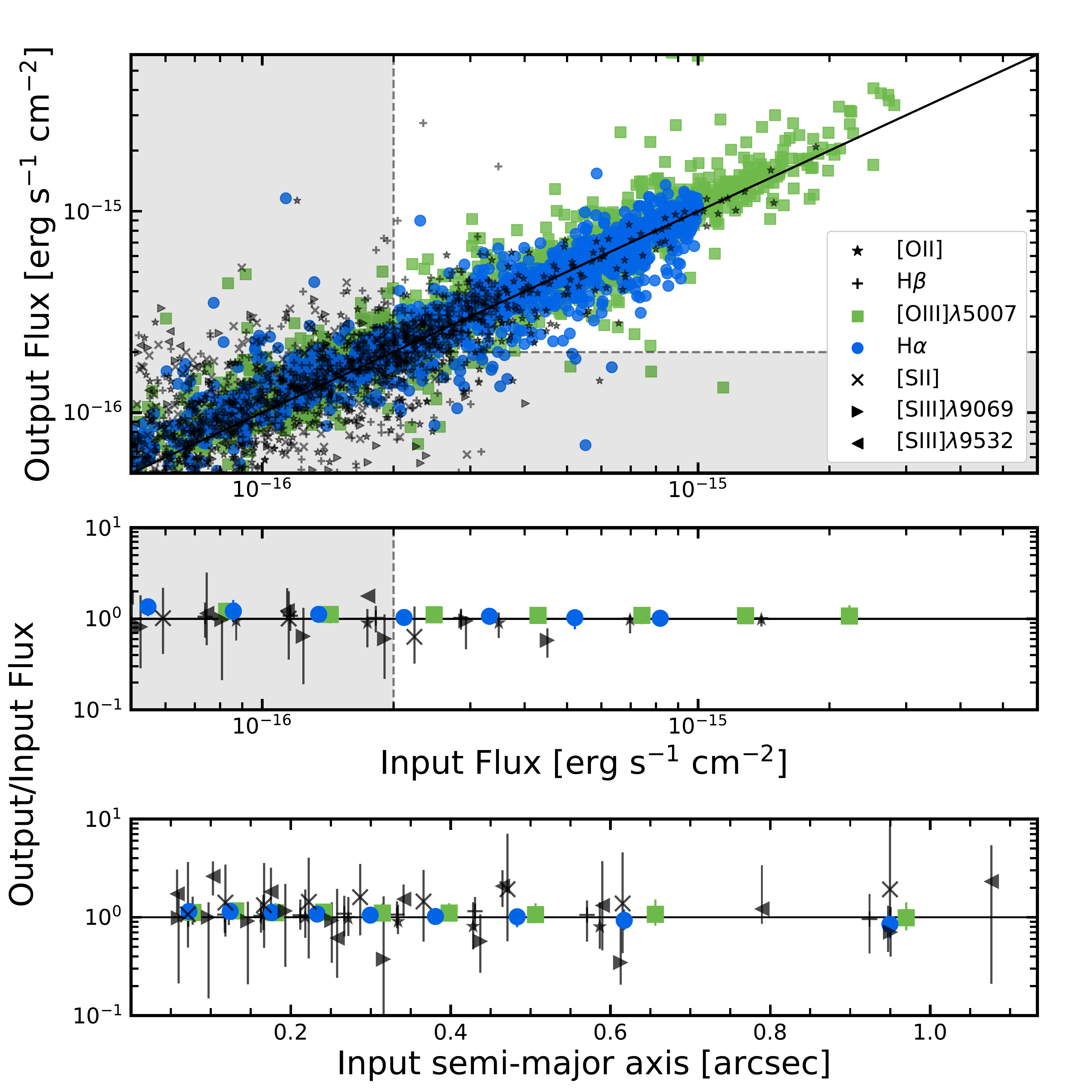}
\caption{
A comparison between the input and
measured (or ``output'') emission line fluxes for the simulated sources
described in Section~\ref{sec:wisp}.
The median flux ratios for each line along with 1$\sigma$ error bars are 
shown in bins of input flux (middle panel) and input source semi-major axis 
(bottom panel).
There is neither a dependence on source size nor on line flux down to the 
$\sim$6$\eseven$ \esc, the flux limit of the Euclid Deep survey. 
The \ha\ and \oiiib\ lines are emphasized because they are the most common 
primary lines. The scatter in the relationship for other lines likely reflects 
the fact that these lines are often measured below the S/N threshold of the 
emission line detection process.
We conclude from the flux agreement for \ha\ and \oiiib\ that we are not 
systematically missing flux in our measured spectra.
\label{fig:fluxcomp}}
\end{figure}
The first case is analogous to slit or fiber losses in spectroscopic 
observations obtained with apertures smaller than the source. 
In this case, flux loss depends partially on the color and morphology
of the galaxies \citep[e.g.,][]{brinchmann2004} and therefore is not a simple 
systematic flux offset. 
We can test the extent of the flux lost in WFC3 slitless 
grism data with the simulations described in Section~\ref{sec:wisp}.
The \texttt{aXeSIM} software generates the synthetic spectrum of a source 
by convolving an imaging template with a template spectrum. We use 
two-dimensional Gaussians to model the sources, but in principle 
any image of the source can be used. Regardless, the shape and size of 
the emission at each wavelength in the synthetic spectrum is assumed to be 
the same as in the direct image. 
The one-dimensional spectrum is then produced by collapsing the extracted 
spectral stamp along the spatial axis. If the extraction window is too small 
in the spatial direction, the flux in the one-dimensional spectrum will 
underestimate the total. 
We can therefore determine what fraction of flux is lost in the emission 
lines by comparing the input values with those recovered and measured by 
the full analysis process. 
The ratio of the measured to input flux is shown in 
Figure~\ref{fig:fluxcomp} as a function of input line flux 
and input semi-major axis, where we find that the fluxes are consistent 
for the primary lines \ha\ and \oiiib\ down to the flux limit of the 
Euclid Deep survey.
Additionally, there is no clear dependence on source size (bottom panel),
indicating that the extraction windows are adequate even for the largest 
sources, i.e., those most likely to have a surface brightness that drops 
below our detection threshold. 
However, we note that due to incompleteness, there are very few sources 
with semi-major axis $a>0\farcs7$.

%%%%%%%%%%%%%%%%%%%%%%%%%%%%%%%%%%%%%%%%
\subsubsection{\han\ Emission Size Measurements} \label{sec:hasize}
\begin{figure*}
\plotone{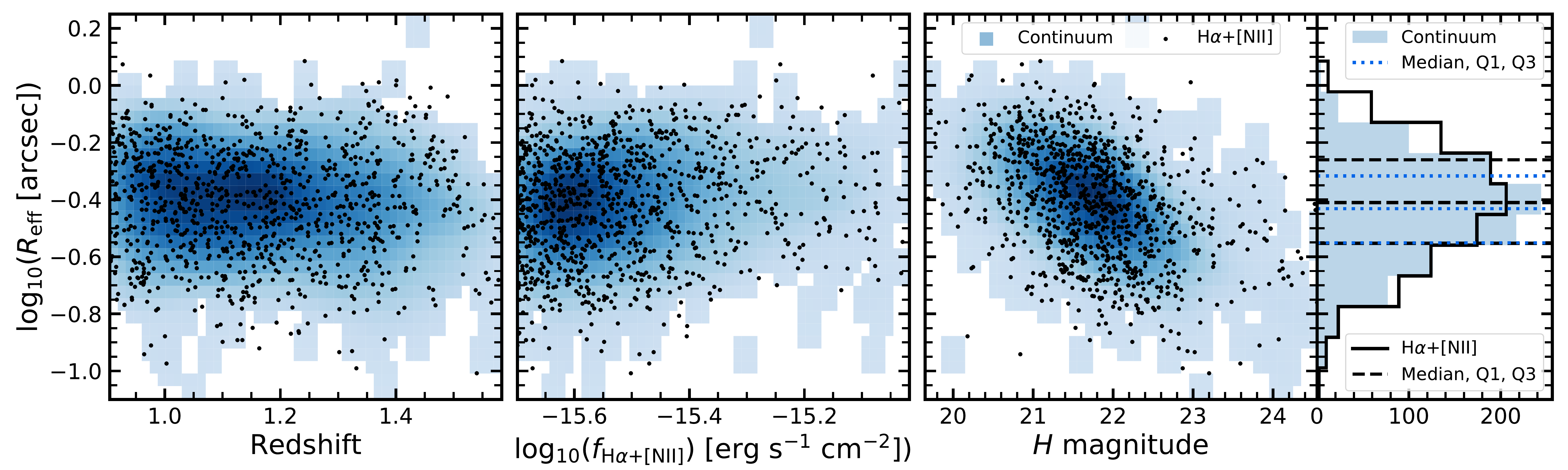}
\caption{The half-light radii (\reff) measured in the continuum 
and \han\ emission line maps as a function of redshift (left), \han\ line 
flux (middle), and $H$ magnitude (right). The continuum sizes are 
represented by a Gaussian kernel density estimation shown in blue, where 
darker colors indicate a larger concentration of sources. The \han\ sizes 
are shown as black points. 
The two size distributions are plotted in the right-most panel 
with blue dotted and black dashed lines indicating the 
median values and interquartile ranges of the continuum and \han\ sizes, 
respectively. The median \reff\ are $\sim$$0\farcs37-0\farcs39$, and for 
reference, the Euclid NISP pixel size is $0\farcs3$.
\label{fig:sizedist}}
\end{figure*}
As discussed in Section~\ref{sec:sizemodels}, we fit 
S\'{e}rsic models to the continuum images and emission line maps.
We show the size distributions as a function of redshift, \han\ flux, 
and $H$ magnitude in Figure~\ref{fig:sizedist}, where \reff\ refers to 
the half-light radius of the S\'{e}rsic profile. We have removed from this 
figure and analysis six 
sources\footnote{The spectra of four of these six sources were 
contaminated by continuum emission from bright neighbors. While this 
contamination did not overlap with the emission lines, it did result in an 
over-subtracted continuum in the \han\ maps. The other two sources were 
very close to detector artifacts in the direct images and therefore had 
incorrectly-measured continuum sizes.} for which the models could not be 
successfully fit ($<1$\% of the sample). 
The \reff\ measured in the \han\ emission line 
maps are shown as black points while the continuum \reff\ are represented 
by the shaded, two-dimensional histogram calculated using a Gaussian 
kernel density estimation. 
The continuum and \han\ size distributions are shown in the right-most panel 
along with the median and quartiles in blue dotted and black dashed lines,
respectively. 
The median continuum \reff\ is $0\farcs37$ ($\sim$2.9$-$3.1 kpc for 
redshifts 0.9$-$1.5) with an interquartile range 
of $0\farcs2$ ($=0\farcs48 - 0\farcs28$). The \han\ size distribution is slightly 
wider with a median of $0\farcs39$ ($\sim$3.0$-$3.3 kpc) and an 
interquartile range of $0\farcs27$ ($=0\farcs55 - 0\farcs28$).

These \reff\ values are on the low end of what is presented for 
continuum emission by \citet{vanderwel2014} and for \ha\ and continuum 
emission by \citet{nelson2016}.
For example, \citet{nelson2016} find $r_{1/2,\mathrm{H}\alpha} = 2.91$ kpc and 
3.10 kpc for galaxies in the mass ranges $9.5<\mathrm{log}(M_*) < 10.0$ and 
$10.0<\mathrm{log}(M_*) < 10.5$, respectively, whereas we expect some if not 
all of the bright ($H\lesssim24$) 
galaxies in the WS to be in a higher stellar mass bin.
The discrepancy between our measurements and those of these 
other works is most likely due to the lower surface brightness limits 
these authors reach by stacking images and spectral stamps
\citep[e.g., $\sim 1\times 10^{-18}$ \esc arcsec$^{-2}$ by][]{nelson2016},
enabling them to recover more of the flux in the 
wings of each source. In this paper, we measured \reff\ on individual stamps
to reflect the role the continuum and emission line sizes have on the 
selection function of slitless spectroscopic surveys, but we note that 
stacking would be required to statistically recover and measure the sizes of 
such sources.

These half-light radii have been deconvolved with 
the empirical WFC3 PSF and therefore represent the intrinsic sizes measured
to the depth of the \hst\ grism observations. 
The median \reff\ measured in both the continuum and \han\ emission is 
$<0\farcs1$ 
larger than the size of one pixel on the 
NISP instrument ($0\farcs3$). Approximately 20\% (40\%) of the WS presented 
here have half-light radii smaller than the NISP pixel in both (either)
the continuum and (or) \han. 
As more flux will be concentrated in each source's central pixel on the 
NISP detector, sources will be more under-sampled in Euclid observations than 
they are when observed by \hst. However, the planned multiple dithers will 
help compensate for the larger pixel scale. 

\begin{figure}
\epsscale{1.1}
\plotone{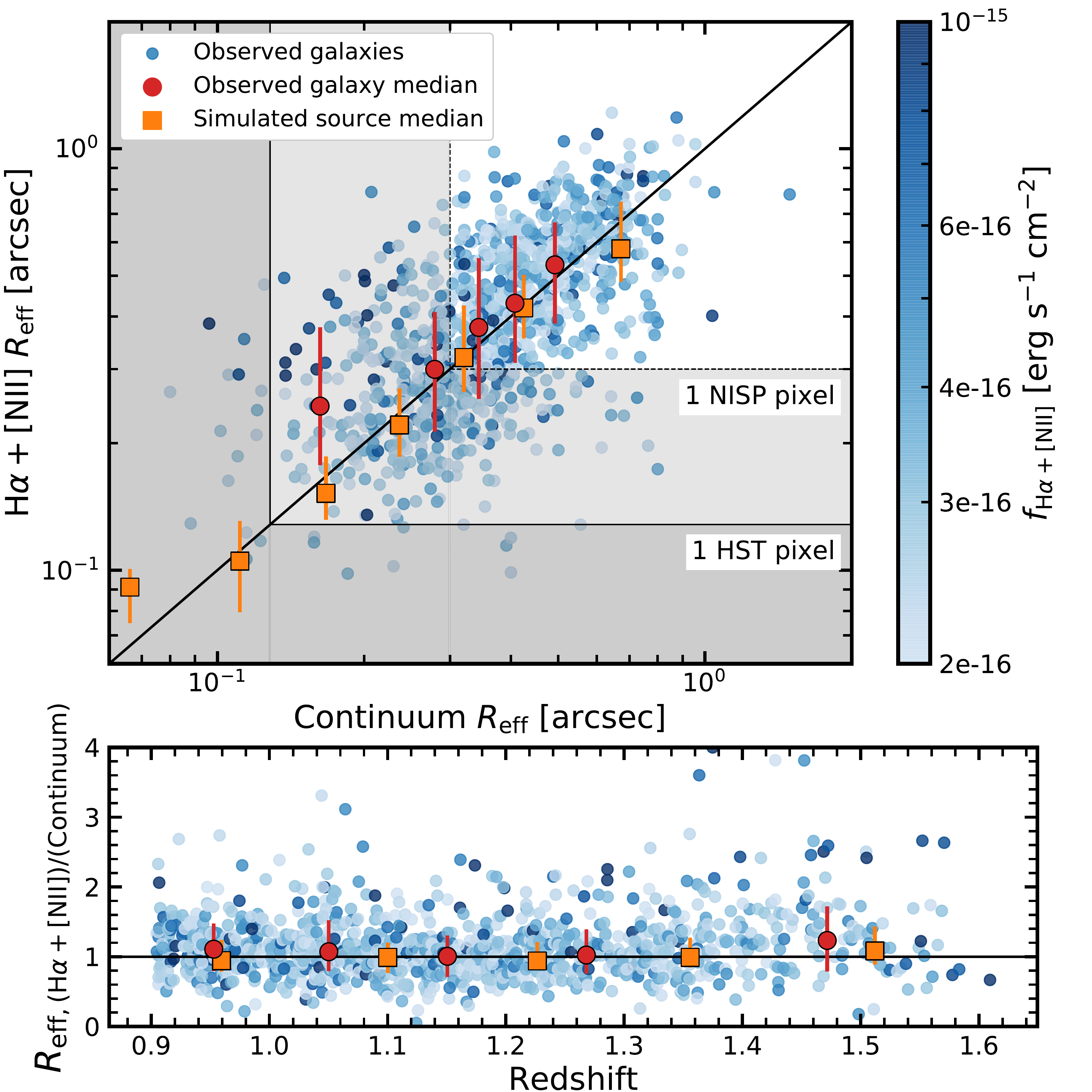}
\caption{
The effective radii of sources measured in both the continuum and \han\
emission line maps, color-coded by \han\ flux (top panel). The ratio of 
$R_{\mathrm{eff,H}\alpha \mathrm{+[NII]}} / R_{\mathrm{eff,continuum}}$
is plotted as a function of redshift in the bottom panel. Though there is 
large scatter, the relation between the two sizes depends on neither line 
flux nor redshift. The red circles show the 
median values and 1$\sigma$ scatter measured for an equal number of sources 
per bin. The median values and scatter calculated for the simulated sources 
from Figure~\ref{fig:simsizes} are shown as orange squares for comparison. 
For reference, the shaded regions indicate the size of 
one WFC3 pixel ($0\farcs13$, darker region) and one Euclid NISP 
pixel ($0\farcs3$, lighter region). 
\label{fig:haniisizes}}
\end{figure}

We next compare the \reff\ measured
in the continuum to that of the \han\ emission in Figure~\ref{fig:haniisizes}. 
The two sizes are correlated, but with significant scatter. The 
standard deviation of the relation between the continuum and emission 
line sizes is $\sim0\farcs2 - 0\farcs35$, compared to the $0\farcs05-0\farcs15$ 
measured for the simulated sources. As the observed and simulated 
data have been fit with different models, we also fit the observations with 
Gaussian models to test whether this increase in scatter is due to model 
choice. However, the Gaussian \reff\ of the observed galaxies 
is very similar to what is shown in Figure~\ref{fig:haniisizes}, and so 
the observed scatter must be due in part to other causes.

Many of the sources above the one-to-one correlation in the left panel of 
Figure~\ref{fig:haniisizes}, where the measured \han\ 
$R_{\mathrm{eff}}$ is larger than 
that of the continuum, have broad \han\ line profiles along the 
dispersion direction. The extent of the emission in these cases is not spatial, and these 
sources are erroneously fit with elongated profiles. 
For others, the \han\ emission is more extended than the continuum as 
discussed by, e.g., \cite{nelson2016}.
For some of the sources with \han\ $R_{\mathrm{eff}}$ smaller than the 
continuum, we may be measuring small knots of emission concentrated within a 
smaller radius than the full galaxy.
In these cases, we may be detecting \ha\ from star-forming clumps 
\citep[e.g.,][]{bournaud2014,zanella2015,mandelker2017}
while any more extended line 
emission has a surface brightness below our detection sensitivity. 
We note that such sources may have increased noise in the extracted, 
one-dimensional spectra, as the extraction windows and extraction weights are 
determined by the size of sources in the direct imaging (see 
Section~\ref{sec:wisp}). The scatter in the \han-continuum \reff\ relation 
in Figure~\ref{fig:haniisizes} can therefore reveal important information 
about the optimization of the spectral extraction process and the resulting 
S/N measured for emission lines.

Part of the scatter may be due to the fact that 
S\'{e}rsic models are simplified representations of galactic light profiles. 
For example, approximately 25\% of the WS sample have either
continuum or line emission characterized by clumps or other structure 
that may indicate merging or interacting systems. 
In these cases, the single-component fits are too simple to properly 
model the emission.
However, much of this sub-structure will be unresolved when observed with 
the larger NISP pixel, and single-component models may provide better fits 
to source continuum and line emission.
On the other hand, most disk galaxies have a bulge 
component that is best fit with larger $n$. More realistic models may be 
achieved by allowing for a two-component model fit consisting of both 
disk-like and bulge-like profiles. 
The resulting measurements could then be compared with the distributions 
of bulge and disk lengths and axis ratios in the Euclid Flagship 
mock catalog. 

Regardless, the distribution of emission sizes measured in slitless data, 
the relation between 
the continuum and line emission, and the observed scatter in this relation
are important quantities for evaluating the effects of the selection 
function of future grism-based galaxy redshift surveys.

%%%%%%%%%%%%%%%%%%%%%%%%%%%%%%%%%%%%%%%%
\subsection{Equivalent Width of \han} \label{sec:ew}
The emission line EW, a measure of the strength of the emission, is a 
very important property of ELGs that must be correctly included in 
forecasts for emission line studies. 
Hydrogen recombination lines such as \ha\ and \hb\ are produced by 
the ionizing radiation from young, massive stars
while the strength of the stellar continuum reflects the buildup of 
emission from the older, less massive population. The EW of \ha\ is therefore 
an estimate of the ratio between the average star formation from current and 
past events. It is a measure of a galaxy's specific star formation rate, or 
the star formation rate per unit stellar mass. Given an assumed 
star formation history, the specific star formation rate 
can be converted to an age for the galaxy. 
The addition of \nii\ complicates this picture, as the \nii\ 
contribution to the \han\ line flux depends on factors such as 
mass, metallicity, star formation rate, ionization parameter, and AGN 
activity and also varies with redshift 
\citep[e.g.,][]{bpt,erb2006,kewley2013,masters2016,kashino2017,faisst2018}.
It is crucial that the simulations created to evaluate the survey design 
of missions such as Euclid reproduce the physical properties, and not 
just the number counts, of the selected population that will be observed. 
As discussed in Section~\ref{sec:counts}, \ha\ and \nii\ will be 
blended at the resolution of the Euclid Red grism 
for all but the most compact sources, and so the observed joint 
\han\ EW distribution should also be reproduced in the simulations.

\begin{figure}
\plotone{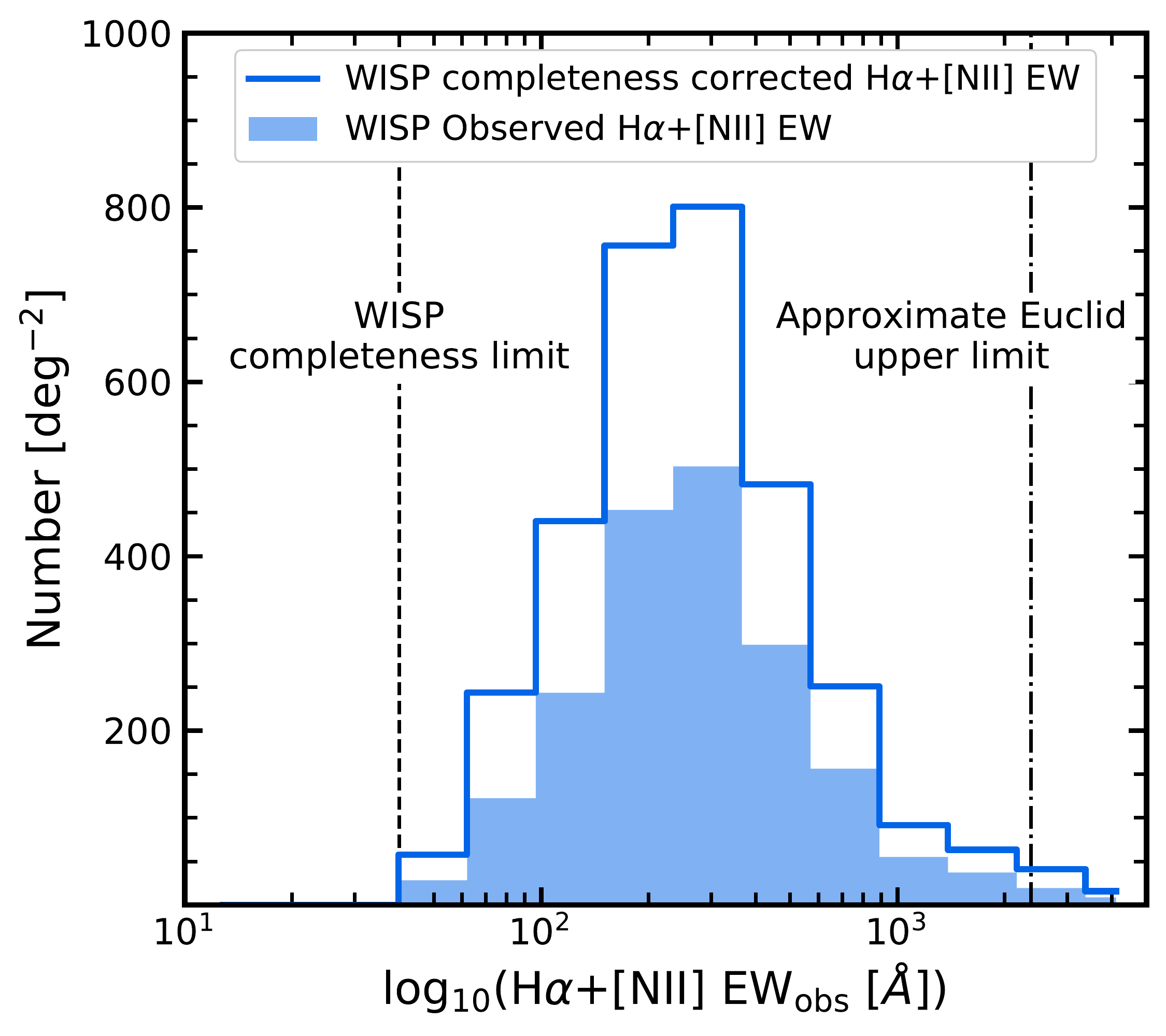}
\caption{The distribution of \han\ EW in the Euclid WS. The median 
EW$_{\mathrm{obs,H}\alpha \mathrm{+[NII]}} \sim 250$~\AA. The distributions
of the observed and completeness-corrected samples are shown as filled and 
empty histograms, respectively.
\label{fig:ew}}
\end{figure}

In Figure~\ref{fig:ew}, we present both the observed and 
completeness-corrected \han\ EW distributions of the Euclid WS.
The median observed \han\ EW is 250~\AA, which corresponds to 
$100-125$~\AA\ for galaxies at $z\sim1-1.5$. The interquartile range of the 
observed EW is $160.2 - 397.5$~\AA. 
The EW distribution ranges from 40~\AA
(the EW completeness limit for the WISP emission line detection algorithm)
to $\gtrsim$4000~\AA. 
The WISP completeness limit represents the EW below which the detection 
algorithm does not reliably detect emission peaks in the grism spectra
(see Section~\ref{sec:wisp}) and is therefore applicable to 
all grism observations run through this software.
The Euclid Wide survey will have an approximate upper EW limit of 
$\sim$2370~\AA, calculated for the $2\times10^{-16}$ \esc\ flux limit 
and a $H=24$ (for $\lambda_{\mathrm{pivot}}=18000$~\AA). 
While there are $\sim$32 sources deg$^{-2}$ in the WS above this 
approximate limit, only $\sim$4 deg$^{-2}$ are also fainter than $H=24$. 
The WISP completeness limit and approximate Euclid Wide Survey limit are 
indicated in Figure~\ref{fig:ew} by dashed and dot-dashed lines, respectively.

%%%%%%%%%%%%%%%%%%%%%%%%%%%%%%%%%%%%%%%%
\subsection{\oiii\ Line Profile} \label{sec:oiii}
As described in Section~\ref{sec:linecatalog}, we assume single emission 
lines in the grism spectra are \ha\ unless the line has noticeable asymmetry 
indicative of the \oiii$+$\hb\ line profile. We now briefly consider whether 
this assumption leads us to selectively identify \oiii\ lines with 
asymmetric profiles. 
\begin{figure}
\epsscale{1.2}
\plotone{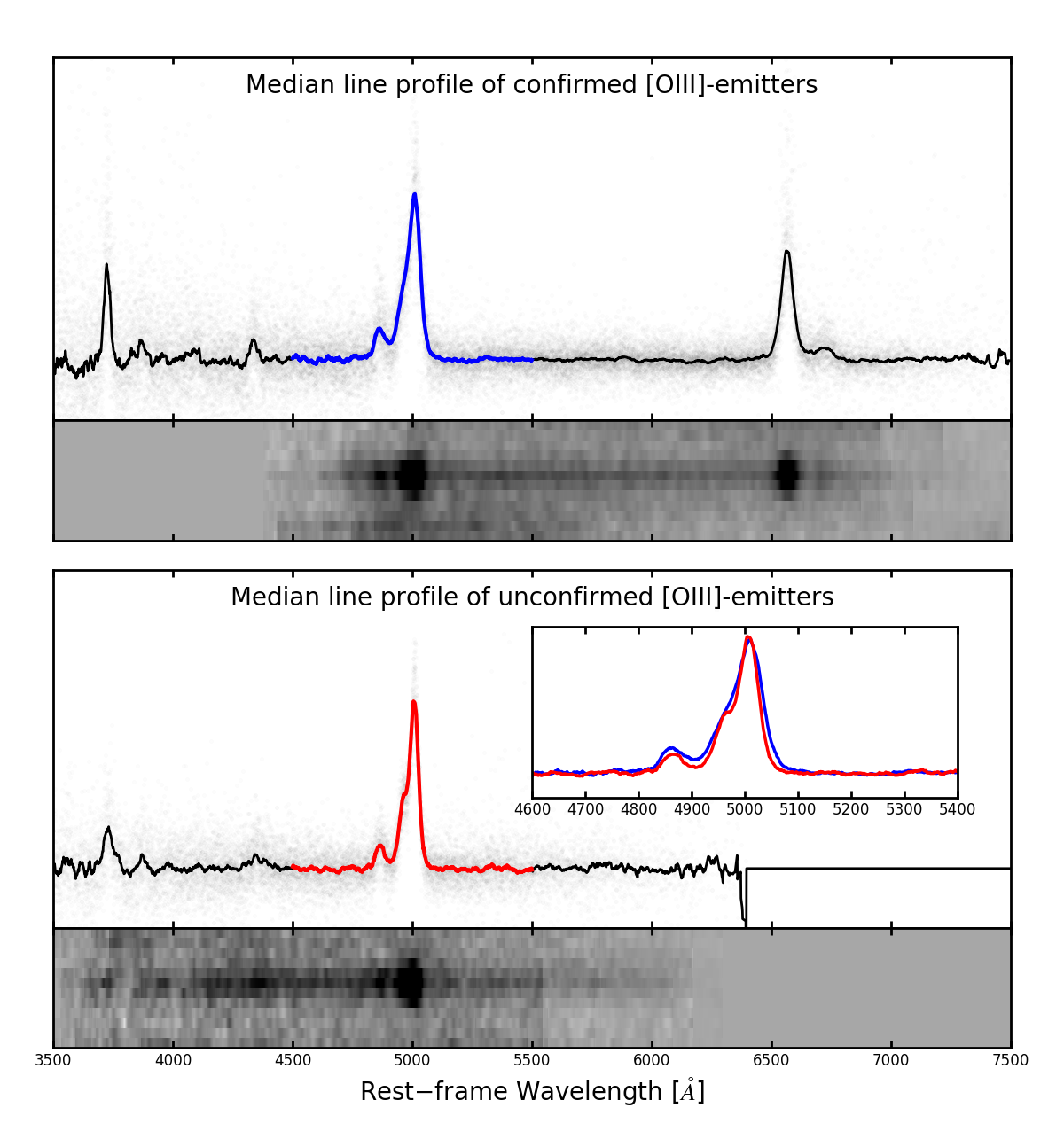}
\caption{
The \oiii\ line profile of sources with secure redshifts (blue, top panel) 
is compared with that from sources with redshifts based on the detection 
of a single emission line (red, bottom panel). The line profile of 
the unconfirmed \oiii-emitters is more asymmetric than that of confirmed 
\oiii-emitters. When fit with a symmetric profile, the residuals 
around the wavelength of the \oiii~$\lambda4959$ line are $> 5\times$ larger 
for the red profile than for the blue.
The asymmetry is needed to make a reliable line 
identification in the absence of additional emission lines.
\label{fig:oiii}}
\end{figure}

We compare the \oiii\ line profiles of sources with multiple lines, and 
therefore secure redshifts, with the profiles of single line emitters that 
have been identified as \oiii. The question is whether a sample of 
sources with multiple lines have, on average, a more symmetric profile 
because the characteristic asymmetry is not needed for line identification. 
We create two subsets of \oiii-emitters randomly sampled from the WISP 
emission line catalog. 
All sources have \oiii\ fluxes with S/N$\geq$5. 
The \han\ fluxes in the first subset also have a S/N$\geq$5, making this a 
sample of confirmed \oiii-emitters.
The second subset is taken at $z\geq1.6$, where \han\ has redshifted out of 
G141, and excludes any sources with an \oii\ S/N$\geq2$. 
We restrict both selections to $z\geq1.24$ so all emission lines are 
measured in G141 with the same resolution and dispersion.
There are $\sim$120 sources in each sample. 
The individual one-dimensional spectra are represented by faint dots in 
Figure~\ref{fig:oiii}, and the median spectrum is shown as the 
black curve. 
We also median combine 20\% of the two-dimensional spectra for each sample, 
displayed below the one-dimensional spectra.
All spectra (one- and two-dimensional) are shifted to the
restframe and normalized by the integrated \oiii\ line flux.

The \oiii\ line profiles for each sample are indicated in blue (confirmed) 
and red (unconfirmed). As can be seen in the inset in the bottom panel, the 
median profile of confirmed \oiii-emitters is indeed more symmetric than that
of the unconfirmed. 
To quantify the level of asymmetry, we fit both line profiles with 
a Gaussian function and measure the residuals. Within $\pm$20~\AA\ of the 
\oiii~$\lambda$4959 line, the residuals of the fit to the red profile 
in Figure~\ref{fig:oiii} are a factor of $>5$ times larger than that of 
the fit to the blue profile.
The median 5$\sigma$ depth in the WISP spectra 
at the wavelengths of the \oiii\ lines is $6\times10^{-17}$ \esc. 
This bias could be even more pronounced for 
shallower data such as that of the Euclid Wide survey, where the 
\oiii$\lambda4959$ line will fall below the detection limit more often than 
in deeper spectra. The resulting \oiii$\lambda5007$ line profiles may 
appear symmetric and be more likely to be identified as \ha\ under visual 
inspection. However, the Euclid Red grism will have a higher spectral resolution
($R\sim380$ compared with $R\sim130$ for G141), and galaxies will appear
more compact on the larger pixel scale ($0\farcs3$ versus $0\farcs13$).
Additionally, as discussed in Section~\ref{sec:3dhst},
unaided human classification will not be a feasible method for line 
identification in the Euclid data.

%%%%%%%%%%%%%%%%%%%%%%%%%%%%%%%%%%%%%%%%
\subsection{Redshift Accuracy} \label{sec:zaccuracy}
The measurement of the BAO signal in galaxy clustering requires accurate 
distance measurements to a large sample of galaxies. It has been
shown through simulations that the redshift accuracy for a survey such as 
Euclid must be $\sigma_z / (1+z) \leq 0.1$\% \citep{wang2010,laureijs2011}. 
As shown by \cite{colbert2013} with simulated sources added to real WISP 
fields, the required redshift accuracy is achievable with $R>200$ grism 
spectroscopy (see their Figure~5).
Here we provide empirical confirmation of the redshift accuracy that can 
be expected from slitless spectroscopy.

\begin{figure}
\epsscale{1.2}
\plotone{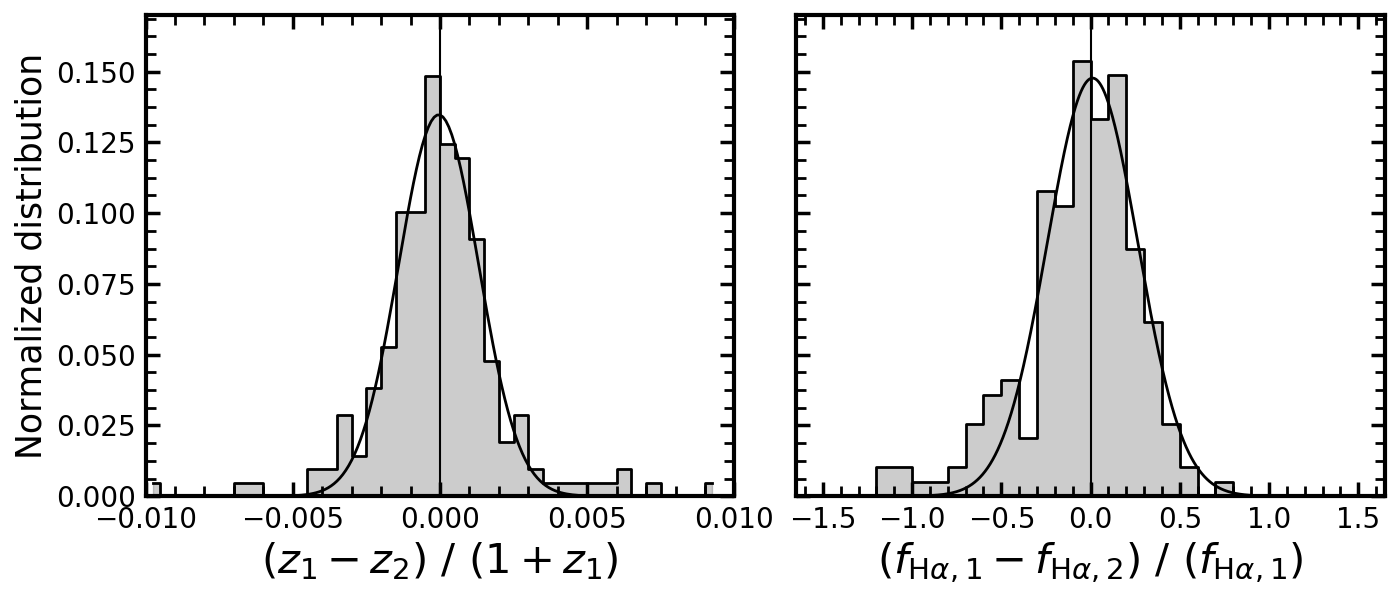}
\caption{
The redshift and \han\ flux accuracies measured empirically from WISP 
data are shown in the left and right panels, respectively. Here we are 
comparing the redshifts and \han\ fluxes from multiple measurements 
of the same set of sources observed in overlapping WISP fields.
\label{fig:delta_real}}
\end{figure}

We perform an empirical measurement of the redshift accuracy using fits 
to the grism spectra of WISP sources that were observed more than once.
Over the six cycles of parallel observations, there are 
36 WISP fields that overlap to some degree with another field. 
There are therefore $\sim$140 sources that have been observed multiple 
times, often with very different exposure times, field depths, and roll 
angles. 
In order to increase the sample size, we consider all possible permutations 
of pairs of observations of a given source. We randomize the order 
in which we calculate the delta redshift to avoid systematic shifts that
may be introduced if a subset of these WISP fields have problems.
Such problems could include issues with the wavelength calibration 
or noisy grism data, which would increase the uncertainty in the 
measured emission line centers. 
In Figure~\ref{fig:delta_real}, we show this empirical measurement 
of the redshift accuracy ($\sigma_z/(1+z)=0.00136$) as well as a similar 
measurement of the accuracy of the \han\ fluxes of these sources. 
Note that the redshift accuracy presented here of $\sim$0.14\% is the 
result of a fit to the difference in two redshift measurements, and 
therefore has twice the variance of either measurement alone. 

In both the simulated data \citep{colbert2013} and the empirical measurement 
presented here, the redshift accuracy measured from the WFC3 slitless 
data is on the order of 0.1\%, indicating the level achievable for 
future grism-based galaxy redshift surveys.
However, such surveys will also have to contend with 
redshift contamination from misidentified emission lines. We quantify the 
expected fraction of contamination as a function of survey flux limit 
in the following section.

%%%%%%%%%%%%%%%%%%%%%%%%%%%%%%%%%%%%%%%%
\subsection{Contaminating Redshifts}\label{sec:contam}
For proper forecasts of dark energy experiments, a critical parameter is the
purity of the measured redshifts of galaxies, which can be quantified by the
fraction of targets with incorrectly identified emission-lines. 
There are two possible sources of contamination: spurious sources 
such as noise peaks incorrectly identified as emission 
features, and real lines that have been misidentified and are therefore 
assumed to be at the wrong redshift.
The first case depends on the method used for line identification.
For example, \cite{colbert2013} find that $\sim$8.5\% of 
emission lines in the first version of the WISP emission line catalog 
are in fact hot pixels, cosmic rays, or other artifacts. Though we have not 
quantified this fraction in the new catalog, the updated procedure using 
a continuous wavelet transform should improve upon this false detection rate
(see Section~\ref{sec:linecatalog}).

To evaluate the contamination from misidentified redshifts, we use 
the CANDELS multi-wavelength observations available for the 3D-HST 
fields and the full wavelength coverage of the G102$+$G141 WISP observations 
to evaluate the purity of spectroscopic redshifts measured with 
grism data.
Figure~\ref{fig:zrange} shows the redshift ranges for 
which multiple lines will be identified in Euclid spectra. 
For many redshifts only one line will be available. 
In addition, depending on the intrinsic \han/\oiiib\ ratio and the
amount of dust extinction, it is likely that  \han\ will still be the only line
detected, even in the redshift range where both \oiiib\ and \han\ are present.
Indeed, only about 10\% of the WS sources in the proper redshift range have 
both \han\ and \oiiib.
When only individual lines are detected, these are operationally identified as
\ha\ unless other information such as the emission line shape or galaxy color
is available.
However, it is possible that a substantial fraction
of these single lines are in fact \oiii\ at $z_{\rm [OIII]}+1=(z_{\rm H\alpha}+1)\, \lambda_{\rm H\alpha}/\lambda_{\rm [OIII]}$.
We aim to constrain the purity of slitless-selected samples
with two complementary approaches: (1) a comparison with 
spectroscopically-confirmed and photometrically determined redshifts, and 
(2) an analysis of the additional secure redshifts made possible by 
increasing the survey wavelength range. 

\begin{figure}
\epsscale{1.2}
\plotone{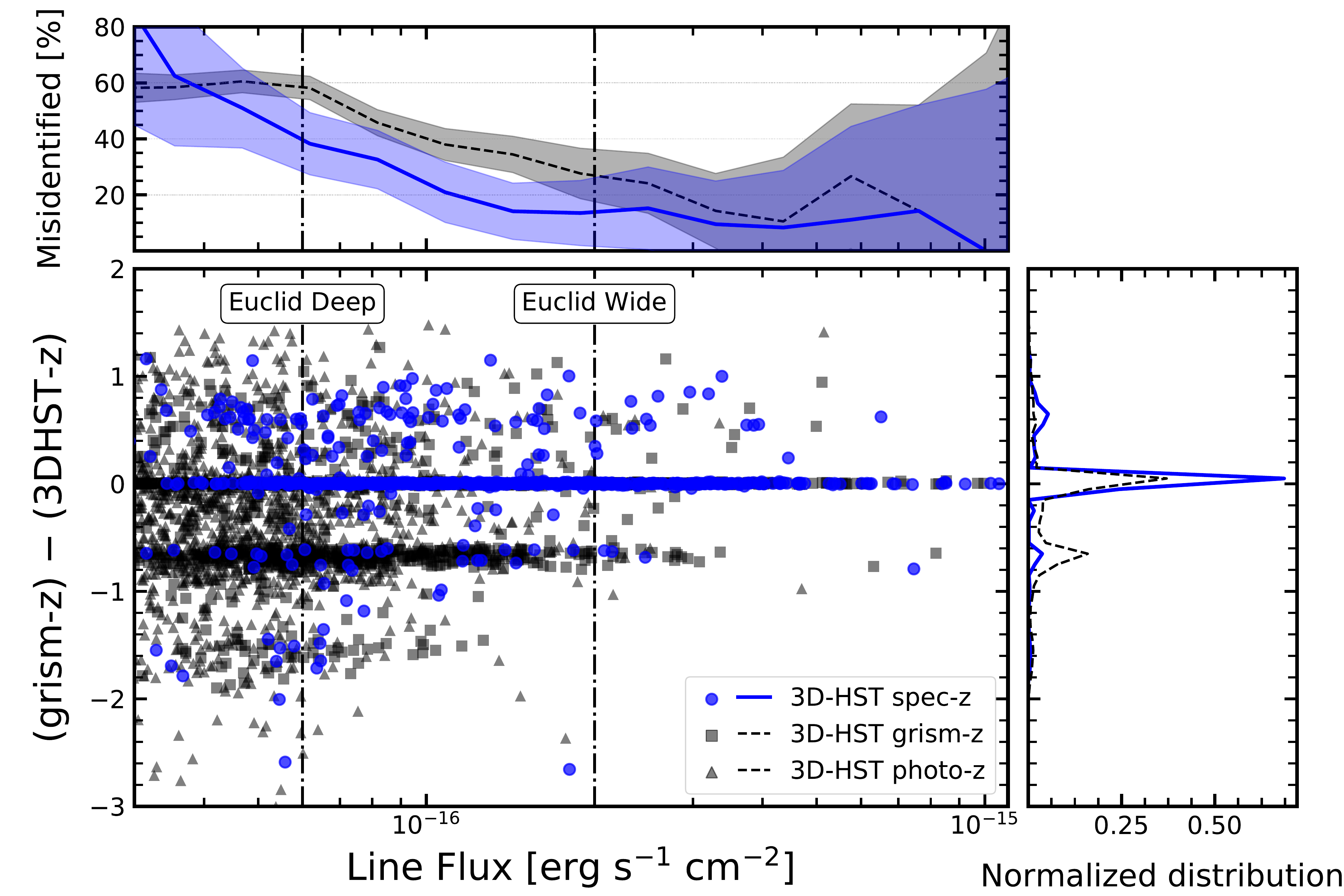}
\caption{
The difference between the grism-identified redshifts used in the WS and 
the redshifts compiled in version 4.1.5 of the 3D-HST catalog for 
single line emitters is shown as a function of line flux. 
All sources were identified as \han-emitters in the grism spectra using the 
emission line detection software discussed in Section~\ref{sec:sample}.
Yet many appear to be lines at a different redshift (typically \oiii) based 
on their spectroscopic redshifts (blue circles), photometric redshifts
(black triangles), or grism redshifts fit using photometric redshifts as a 
prior (black squares). This redshift 
misidentification rate depends on the survey flux limit. 
Approximately 14\% (40\%)  of the single line emitters plotted in blue are 
misidentified at the depth of the Euclid Wide (Deep) Survey.
\label{fig:misid}}
\end{figure}

First, we use the redshifts compiled in version 4.1.5 of the 3D-HST 
catalog\footnote{\url{http://3dhst.research.yale.edu/Data.php}} to 
determine the fraction of 3D-HST grism redshifts assigned through the 
WISP emission line detection procedure that have been 
misidentified. In Figure~\ref{fig:misid}, we explore this redshift 
misidentification as a function of line flux using the best available 
redshift for each source (``z\_best'') from the 3D-HST catalog. 
Sources where ``z\_best'' is a spectroscopic redshift from 
\cite{skelton2014} are shown as blue circles. All other redshifts 
are determined either from grism measurements (black squares) or from 
spectral energy distribution fitting using the full suite of available
CANDELS photometry (black triangles). 
Note that here the grism redshifts are those from the 3D-HST data release,
which include the CANDELS photometric redshifts as a prior,
rather than the measurements performed using the WISP emission line 
detection described in Section~\ref{sec:3dhst}.
We consider all \han-emitters in the 3D-HST catalog 
where the redshift is based on a single line. 
We select galaxies with a ``z\_best''~$\geq$~0.9 to match the 
WS selection. 

The prevalence of misidentified single emission lines 
depends on the survey depth. At the depth of the Euclid Wide Survey 
$\sim$14\% of the single line emitters with spectroscopic 
redshifts (blue circles) assumed to be \han\ are in fact a different 
emission line. This percentage increases to $\sim$40\% at the depth of the 
Euclid Deep Survey, below which the sample size of sources confirmed 
spectroscopically decreases. 
As it is prohibitively difficult to follow-up every grism detection of an 
ELG from the ground, we also show the misidentification percentage for 
the sources with no slit-based spectroscopic redshift, where $\sim$25\% to 
$\sim$60\% 
of sources with single emission lines are misidentified. 
For the majority of the Euclid Wide Survey, photometry --- and efforts to 
calibrate photometric redshifts \citep[e.g.,][]{masters2015} --- will be 
critical for correctly identifying single emission lines in the grism spectra
and improving the sample purity attained in the Wide Survey. 

Measuring the BAO signal from galaxy clustering measurements requires a 
full understanding of the sample redshift contamination, as galaxies 
with misidentified redshifts will reduce the strength of the clustering 
signal. 
The Euclid Deep Survey will therefore provide a redshift calibration 
sample, which will be used to quantify the contamination fraction present 
in the Wide Survey. For this calibration effort, the Deep Survey will aim 
to achieve a purity of 
$p>99$\% over the 40 deg$^2$ region, where $p$ is 
the number of sources with correctly identified redshifts divided by the 
total number measured \citep{laureijs2011},
$p=N_{\mathrm{correct}} / N_{\mathrm{measured}}$.
We use the WISP emission line catalog to estimate 
the purity of the emission line sample observed in the Euclid Deep Survey.
From the full WISP emission line catalog (Section~\ref{sec:linecatalog}),
we only consider fields with spectral coverage in both the G102 and the G141 
grisms. Additionally, we include in the following analysis only emission line 
galaxies with secure redshifts (i.e., measured with at least two emission 
lines). Having observations in both grisms ensures a spectral coverage 
between $0.85 \leq \lambda \leq 1.65 \mu$m. Given the emission lines 
considered in the redshift determination (\oii, \hg, \hb, \oiii, \han, 
\sii, \siii$\lambda$9069, \siii$\lambda$9532, and \hei$\lambda$10830), the 
catalog derived for the fields with both grisms spans the redshift range 
between $z\sim0.25$ and $z\sim2.3$. 
The approximate $5\sigma$ depth of the selected WISP 
fields is $5\times 10^{-17}$ \esc, consistent with the expected line flux limit 
in the Euclid Deep Survey observations with the Red grism.
\begin{figure}
\epsscale{1.2}
\plotone{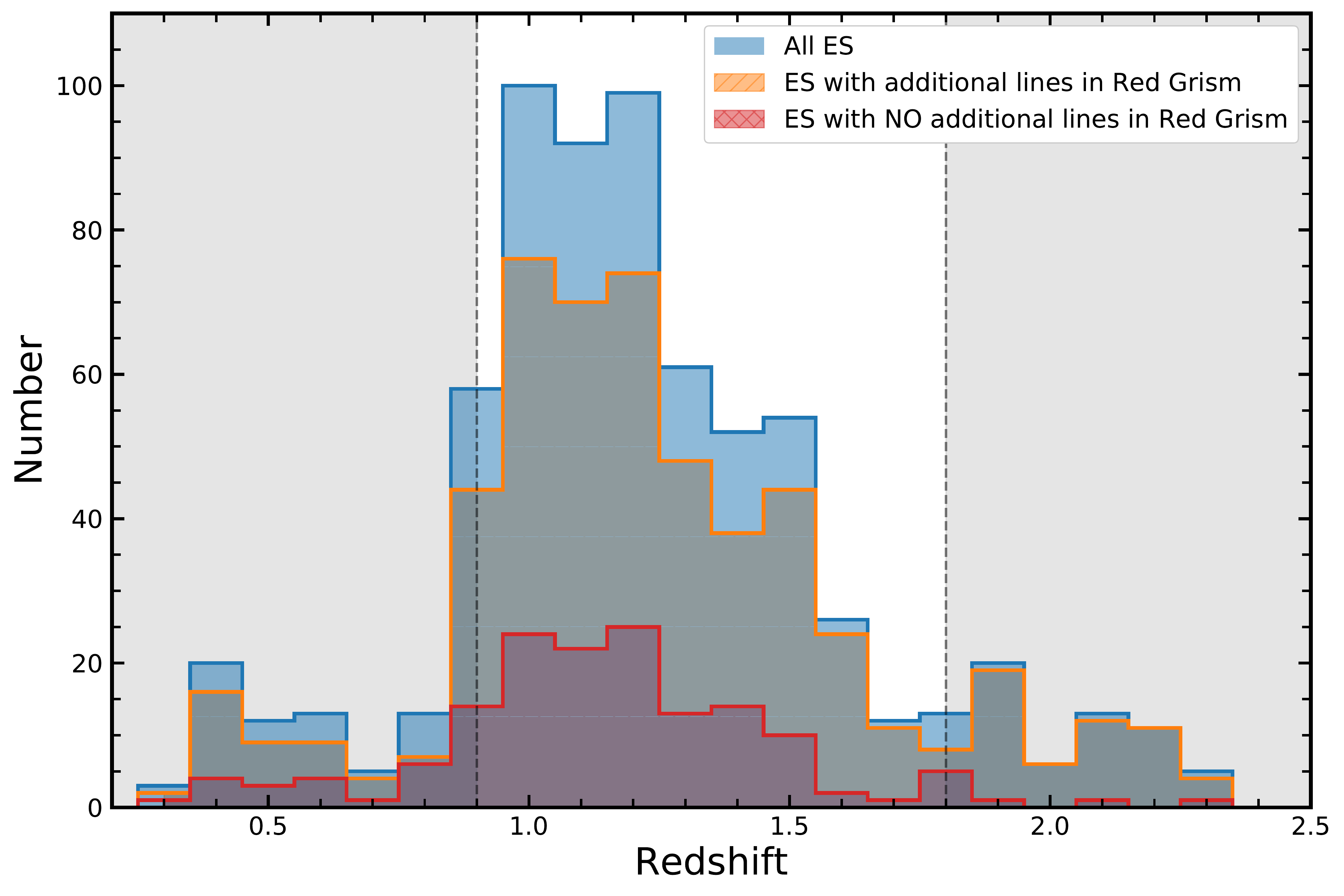}
\caption{
Redshift distribution of galaxies in the Euclid Shallow sample, defined as 
all WISP sources with at least one emission line with 
$\lambda_{\mathrm{line}}>1.25 \ \mu$m and 
$f_{\mathrm{line}}> 2\times 10^{-16}$ \esc, regardless of their redshifts 
(blue histogram). The orange and red histograms split the sample into 
objects with and without additional emission lines in the Euclid Red grism 
at the depth of the Deep Survey. The objects in the red histogram that fall 
in the shaded region, those with $z<0.9$ or $z>1.8$, would result in 
incorrect redshift determinations and 
correspond to about 6\% of the ES sample.
\label{fig:purity}}
\end{figure}

From this two-grism WISP catalog, we apply the same selection criteria 
described in Section~\ref{sec:selection} to create a sample 
analogous to that which will be selected using the NISP Red Euclid grism. 
We call this sample the Euclid Shallow (ES) sample, and it is the same 
as the WS of Section~\ref{sec:selection} but only includes WISP sources 
from fields observed with both grisms. 
The redshift distribution of the ES sample is shown in 
Figure~\ref{fig:purity} (blue histogram). In the WISP catalog, sources 
with redshift in the $0.4<z<1.6$ are mostly selected via their \ha\ emission 
line. At redshifts $z\gtrsim1.6$, galaxies are selected via the \oiii\ 
emission lines, while at redshifts $z\lesssim 0.4$, galaxies are identified 
because of the \siii\ and \hei\ lines. We stress that the WISP sample 
considered in this analysis only includes secure redshifts measured with two 
or more emission lines.

Multiple emission lines are required in order to be able to perform an 
unambiguous redshift identification. Thus, we look at the fraction of 
galaxies in the ES sample that would show additional lines in the wavelength 
range of the Euclid Red grism, with $f_{\rm line}^{add} > 2\sigma$ 
(where $\sigma = 1.4\times 10^{-17}$ \esc\ is the required spectroscopic 
depth of the Euclid deep survey). We find that 77\% of the ES sample have 
multiple emission lines in the  wavelength range of the Euclid Red grism and 
at the depth of the Deep Survey. The redshift distribution for this 
population is shown in Figure~\ref{fig:purity} (orange histogram). 
The red histogram in Figure~\ref{fig:purity} shows the redshift 
distribution of the remaining 23\% of galaxies in the ES sample that would 
be single line emitters in the Euclid Red grism even at the depth of the
Deep Survey.

For single line emitters the simplest assumption is that the line is \ha. 
The red histogram, however, clearly shows that this assumption would get 
the redshift wrong for single-line emitters at $z<0.9$ and 
$z>1.8$. Here we are assuming that Euclid would detect \ha\ for the 
\oiii-selected WISP sources in the  $1.6<z<1.8$ redshift range.
We find that 6\% of all ES galaxies would have incorrect redshift measurements
in a Deep Survey observed with only the Euclid Red grism, corresponding to a 
sample 
purity of 94\%. These incorrect redshifts would make quantifying the 
redshift contamination and purity of the Euclid Wide Sample more challenging.

Given the wavelength range of the G102 grism, we can quantify the extent to 
which the addition of the Euclid Blue grism would improve the sample purity.
The Euclid Blue grism extends the survey wavelength coverage down to 
0.92~\micron. Encouragingly, almost all of the 6\% of objects with 
misidentified redshifts have additional detectable emission lines in this
extended blue wavelength coverage. 
Specifically, 85\% of the misidentified redshifts would be removed, bringing 
the purity of the ES sample up to 99.1\%. The Blue grism would be a valuable 
addition to the Euclid Deep Survey and would allow for a better understanding 
of the fraction of single-line emitters with incorrect redshifts in the 
Wide Survey.

%%%%%%%%%%%%%%%%%%%%%%%%%%%%%%%%%%%%%%%%
\section{Summary} \label{sec:summary}
Upcoming galaxy redshift surveys such as ESA's Euclid mission and NASA's \rst\
mission will use \ha\ and \oiii-selected galaxies to trace the large scale
structure at redshifts of $z \sim 1-2.5$, aiming to understand the nature 
of the accelerated expansion of the universe. 
The constraining power of such surveys is limited by the number density of 
galaxies detected in the survey volumes as well as the redshift accuracy of 
the resulting samples. Additionally, as slitless grism surveys, their 
samples will be the result of  
complex selection functions that depend not only on redshift, but also 
on line S/N, EW, and galaxy size and shape in both the continuum and 
emission lines. The wavelength coverage and resolution of the \hst\ infrared 
grisms provide the valuable opportunity to evaluate the expected selection 
functions of future galaxy redshift surveys 
and their effects on the requirements of the dark energy missions. 

In this paper we create a sample of emission line galaxies 
from the \hst\ programs WISP and 3D-HST$+$AGHAST and explore aspects of 
the sample to present predictions for the Euclid Wide Survey.
The grism data cover 0.56 deg$^2$, approximately equal to the NISP 
field of view. We apply a selection function to match that expected for 
the Euclid Wide Survey, requiring emission line fluxes 
$\geq 2\times10^{-16}$ \esc\ and observed wavelengths $\geq 1.25$ \micron\ in 
addition to S/N and EW cuts necessitated by 
the completeness of the slitless WFC3 data.
We find $\sim$3270 \han-emitters deg$^{-2}$ from $0.9 \leq z \leq 1.6$ and 
$\sim$440 \oiiib-emitters deg$^{-2}$ from $1.5 \leq z \leq 2.3$ in the WS, 
where these number densities have been corrected for the incompleteness of the 
WFC3 grism data. 
The observed number counts are in agreement with predictive models from 
works in the literature, including Model 3 from \cite{pozzetti2016} that has 
been used to calibrate the Euclid Flagship Mock catalog.

We next measure the size and EW distributions for all \han-selected galaxies 
in the WS. As the extraction of spectra in slitless data depends on the 
location, size, and concentration of the sources in direct imaging, it is 
crucial that we understand the relationship between the size of galaxies 
in the continuum and the emission line of interest. 
We 
fit the galaxies in the $H$ band images and \han\ emission line 
maps with S\'{e}rsic profiles, deconvolved with an empirically-determined,
wavelength-dependent PSF. 
The median half-light radii of the galaxies in the continuum and \han\
emission are
$R_{\mathrm{eff,cont}}=0\farcs37$ and 
$R_{\mathrm{eff,H}\alpha \mathrm{+[NII]}}=0\farcs39$, respectively. 
The sizes of the continuum and emission lines 
are correlated, but with significant scatter 
($\sigma \sim 0\farcs2 - 0\farcs35$). 
The median \han\ EW in the observed frame is EW$_{\mathrm{obs,H}\alpha \mathrm{+[NII]}}\sim250$~\AA.
These distributions reflect the properties of the galaxy population 
accessible to redshift surveys performed with slitless spectroscopy, 
and are therefore important 
quantities to include in mock catalogs used to test survey strategies.

Finally, we use the full depth of the emission line catalogs to quantify the 
redshift accuracy and contamination that can be expected for Euclid. Using 
overlapping WFC3 fields where the same sources are observed multiple times,
we measure a redshift accuracy of $\sigma_z/(1+z) = 0.0014$, 
indicative of that which can be achieved by $R\sim200$ slitless spectroscopy. 
We then explore the effect of redshift contamination 
from misidentified emission lines if all single lines are assumed 
to be \ha. By comparing the grism redshifts with the spectroscopic and 
photometric redshifts from the CANDELS catalogs, we find that at the depth 
of the Euclid Wide Survey, $\sim$14-20\% of the resulting sample is likely 
to be incorrectly identified. 
As the majority of galaxies Euclid will detect in the Wide Survey 
will have only one emission 
line in the NISP Red grism wavelength range, it is very important to 
properly quantify this type of redshift contamination. 
The Euclid Deep Survey will be used to calibrate 
the Wide Survey observations and to quantify the redshift contamination rate.
We additionally show that even at the depth of the Deep Survey, approximately 
6\% of emission line galaxies could still be misidentified in the Red 
grism wavelength range. 
However, the addition of the Blue grism to the Deep Survey calibrations 
would significantly reduce the redshift misidentifications and allow for 
a more complete assessment of the Wide Survey redshift contamination. 

The predictions presented in this paper are specific to ESA's Euclid galaxy 
redshift survey as part of the dark energy mission, yet these observations 
can be used as a valuable testbed for other grism-based surveys such as 
\rst\ or for preparations for the NIRCam grism on \textit{JWST}.

%%%%%%%%%%%%%%%%%%%%%%%%%%%%%%%%%%%%%%%%
\acknowledgments
We thank the anonymous referee for a careful review 
and for helpful comments that improved this paper.
We would also like to thank Karlen Shahinyan, Ben Sunnquist, Marc Rafelski, 
Y. Sofia Dai, and Melanie Beck for help with the visual inspection and 
identification of emission lines. 
We would also like to thank Ginevra Favole for her contributions to the
WISP catalog completeness analysis, and Herv\'{e} Aussel for helpful 
discussions regarding the Euclid Science Performance Verification 
efforts.
MBB would like to thank Francesco Valentino for his generosity in 
providing number counts and discussing his analysis. 
% Euclid:
This research was partially supported by NASA ROSES grant 12-EUCLID12-0004.
% Yun Wang support
YW was supported in part by NASA grant 15-WFIRST15-0008 Cosmology with the 
High Latitude Survey WFIRST Science Investigation Team.
% Andrea Cimatti and Lucia Pozzetti support
AC and LP acknowledge the support from the grants PRIN-MIUR 
2015 and ASI n.2018-23-HH.0.
% WISP:
Support for WISP (HST Programs GO-11696, 12283, 12568, 12902, 13517, 13352, and 
14178) was provided by NASA through grants from the Space Telescope Science 
Institute, which is operated by the Association of Universities for Research 
in Astronomy, Inc., under NASA contract NAS5-26555.
% 3D-HST:
This work is partially based on observations taken by the 3D-HST Treasury 
Program (GO 12177 and 12328) with the NASA/ESA HST, which is operated by the 
Association of Universities for Research in Astronomy, Inc., under NASA 
contract NAS5-26555.

%%%%%%%%%%%%%%%%%%%%%%%%%%%%%%%%%%%%%%%%
\vspace{5mm}
\facilities{HST (WFC3)}

\software{Astropy \citep{astropy},
          aXe \citep{aXe},
          aXeSIM \citep{aXeSim},
          SciPy \citep{scipy},
          Source Extractor \citep{bertin1996},
          \galfit\ \citep{peng2010}
          }

%%%%%%%%%%%%%%%%%%%%%%%%%%%%%%%%%%%%%%%%
\appendix

\section{Completeness corrections}\label{app:completeness}

As mentioned in Section~\ref{sec:completeness}, we use two sets of 
completeness corrections in our analysis, one each for the WISP and \tdhst\ 
emission line catalogs. In this way we ensure that the corrections 
are derived using the same procedure as the catalogs to which they are applied. 
In both cases, the completeness is calculated 
with sets of simulated sources that are added to real images and processed 
identically as the real data. We discuss the derivation of the completeness 
corrections for both catalogs in the following sections and compare the two 
line finding algorithms in Section~\ref{app:comp}.

\subsection{WISP Completeness Corrections}\label{app:wispcorrections}
In order to assess the completeness of the WISP Survey and the line 
finding procedure, we create a simulated catalog of 10000 synthetic sources and 
their spectra. The parameters for each source --- redshift, source size 
and shape, \ha\ line flux, \ha\ EW, and \ha/\oiiib\ line 
ratio --- are pulled randomly from uniform distributions chosen to 
bracket the observed values in the emission line catalog. The one 
exception is the \ha/\oiiib\ line ratio, which is pulled from a Gaussian 
distribution centered at the measured ratio for real sources but with a 
standard deviation two times larger. We create synthetic direct 
and grism images of each source using \texttt{aXeSIM} \citep{aXeSim}.
The source images are elliptical Gaussians, which \texttt{aXeSIM} 
convolves with input template spectra to create the dispersed grism 
stamps. The simulated sources therefore have the same size and shape
at all wavelengths. We insert the simulated 
sources into real WISP images from 20 fields, chosen to cover the 
range of exposure times, depths, and filter coverage that exist in a 
survey comprised of parallel opportunities of varied length. We produce 
20 realizations of each field, with 25 simulated sources per realization.
We then process all fields through the full WISP 
reduction pipeline and emission line detection software. 

The completeness is calculated in bins of emission line EW and
``scaled flux'', or the line flux scaled by the grism sensitivity 
in that particular field at the wavelength of the line. This scaled flux is a tracer for line S/N, but also reflects the effect of the varying depths reached in each field of a parallel survey such as WISP.
The bin edges are determined by the distribution of sources in the real 
WISP emission line catalog such that there are an approximately equal number 
of real sources in each bin. The one exception is the bin of lowest EW, which 
we add in order to probe an area of the parameter space with low 
completeness \citep[EW$_{\mathrm{obs}} < 40$~\AA, see][]{colbert2013}.
We note that source size and shape can strongly affect the 
completeness of both the imaging and emission line catalogs, as extended, 
low surface brightness sources may fall below the
adopted \se\ detection thresholds and their low emission line EWs may be
missed by the line detection algorithm. However, the large sources that 
suffer from the highest levels of incompleteness (with semi-major
axis $a \geq 0\farcs7$), 
constitute less than 1\% of either the imaging or the emission line catalogs.
We therefore weight the distribution of input sources by the distribution
of observed sizes in the emission line catalog, allowing us to account for
the effect of source size while considering a two-parameter completeness
correction. 
We use a radial basis function to interpolate the recovery fractions 
across all bins, smoothing over sharp jumps in the completeness at bin edges.
Uncertainties on the completeness corrections are taken as 
$1/\sqrt{N_{\mathrm{rec}}}$, where $N_{\mathrm{rec}}$ is the number of 
recovered sources in each bin of EW and scaled flux.
Completeness corrections are calculated using the line with the 
highest ``scaled flux'' 
in each spectrum as a proxy for the ``primary'' lines identified by the 
detection software. Accordingly, these completeness corrections are applied 
to each source in the emission line catalog -- given the flux
and EW of its strongest line -- rather than to individual emission lines.

\begin{figure}
\plotone{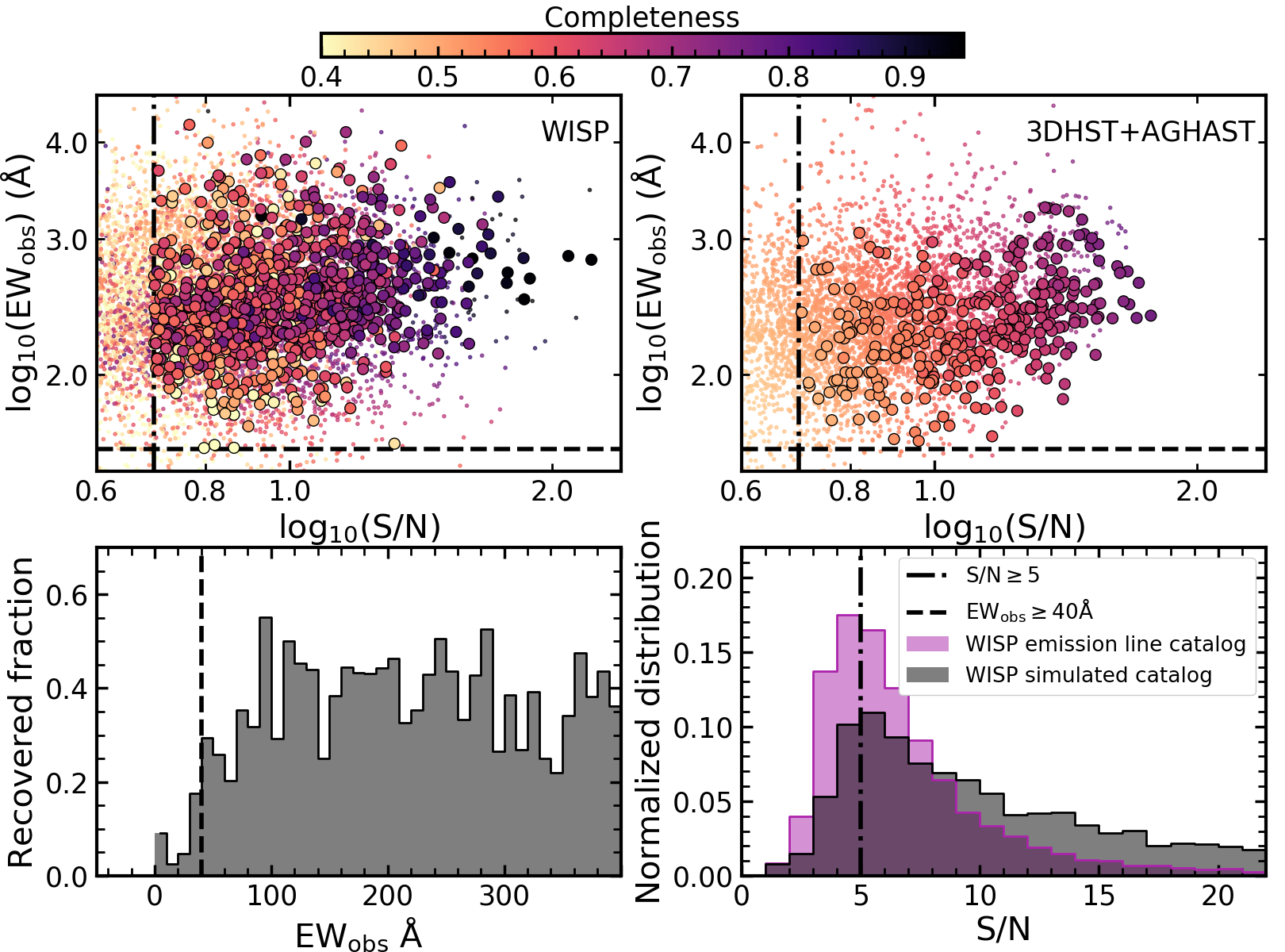}
\caption{A comparison between the completeness corrections applied to the 
WISP (top left) and \tdhst\ (top right) catalogs as a function of emission 
line S/N and observed EW.
In both panels, the larger circles indicate the sources selected as part of
the WS, and the smaller points indicate sources from the full emission line 
catalogs. 
As the corrections for the new method are calculated using a scaled flux
rather than a line S/N, the transition from low to high completeness does not
progress as smoothly in the left panel as it does in the right.
In the bottom panels, we use the simulated sources from the updated 
completeness analysis (grey distributions) to 
demonstrate the EW$_{\mathrm{obs}}$ and S/N thresholds we adopt as part of 
our sample selection. In the bottom left panel, the recovery of simulated 
sources drops rapidly for EW$_{\mathrm{obs}} < 40$~\AA. In the right panel, 
the number of sources in both the real (purple) and simulated catalogs drops
off for line S/N$<$5. 
For all panels, the adopted S/N and EW$_{\mathrm{obs}}$ thresholds are 
shown as dot-dashed and dashed lines, respectively. 
\label{fig:compcompare}}
\end{figure}

In the top row of Figure~\ref{fig:compcompare}, we compare the completeness 
corrections from this new line finding method and that of \citet{colbert2013}.
The completeness corrections 
for sources identified with the new method (WISP catalog) are shown on the 
left, and those for the \tdhst\ catalog using the method described by 
\citet{colbert2013} are displayed on the right. In both
cases we plot the corrections as a function of line S/N and observed EW to 
allow for a direct comparison. 

Finally, this analysis identified two thresholds below which the completeness 
corrections are highly uncertain: line EW$_{\mathrm{obs}} < 40$~\AA\ 
and S/N$<$5, identified in Figure~\ref{fig:compcompare} by dashed and 
dot-dashed lines, respectively.
In the bottom left panel of Figure~\ref{fig:compcompare}, we show the fraction 
of simulated sources that were recovered by the automatic detection algorithm 
binned by observed EW. The fraction drops rapidly for 
EW$_{\mathrm{obs}} <40$~\AA, indicating that 40~\AA\ is an effective 
limiting EW for which the automatic algorithm can detect a significant 
peak above the continuum in the WISP spectra.
Overall, fewer than 10\% of input emission lines with 
EW$_{\mathrm{obs}} < 40$~\AA\ were recovered by the automatic detection 
algorithm, while this fraction increases to 28\% for 
$40 < \mathrm{EW}_{\mathrm{obs}} < 60$~\AA.

In the bottom right panel of Figure~\ref{fig:compcompare}, we show the 
distribution of emission line S/N as measured in both the simulated 
(grey distribution) and real (purple distribution) WISP emission line catalogs.
There is a decrease in the number of sources in both catalogs where the 
strongest emission line present in the spectrum is detected with S/N$<$5.
We have included in the simulated catalog emission lines with fluxes pulled
from a uniform distribution reaching well below the flux limit of the survey.
The bottom right panel of Figure~\ref{fig:compcompare} therefore demonstrates
that we are not complete to sources with emission lines below this cutoff.
Additionally, during visual inspection, reviewers 
are less consistent in their treatment of S/N$<5$ emission lines. 
The fraction of lines accepted by only one reviewer, i.e., when the reviewers 
do not agree that the emission line candidate is real, doubles for lines 
with S/N$<5$. 
We note that these same EW and S/N thresholds were adopted by 
\cite{colbert2013} in their completeness analysis that is used for
the \tdhst\ emission line catalog described in Section~\ref{sec:3dhst}, 
where the motivation is similar.
We therefore include these thresholds in our sample selection criteria 
described in Section~\ref{sec:selection}.

\subsection{\tdhst}\label{app:tdhstcorrections}
\cite{colbert2013} derive completeness corrections for the earlier version 
of the WISP emission line detection process following a similar procedure, 
by adding simulated sources to WISP images and processing them all the way 
through the visual inspection stage. We briefly summarize the steps here 
and refer the reader to \citet{colbert2013} for more details.

A total of 923 model ELGs are generated pulling parameters such as redshift, 
source size, emission line flux, and EW randomly from distributions informed 
by real measurements in WISP data. These are added to 74 realizations of 
WISP fields, with 10-20 simulated sources per field. The automatic line 
detection algorithm identifies sets of 3 or more contiguous pixels above the 
continuum that each have a S/N~$>\sqrt{3}$ (or 2 contiguous pixels that 
are each at a S/N~$>\sqrt{5}$ to account for unresolved objects). Each 
emission line candidate is then inspected by two reviewers, where the 
criteria for accepting or rejecting a candidate is the same as in the 
updated line finding procedure. The reviewers identify each emission line, 
thereby assigning a redshift, and measure line properties by fitting 
Gaussians to the line profiles.  
The completeness corrections of \citet{colbert2013} are calculated in bins 
of line S/N and observed EW, with input distributions weighted by 
source size as described in Section~\ref{app:wispcorrections}. These 
corrections are applied to the \tdhst\ catalog in the top right panel of 
Figure~\ref{fig:compcompare}.

\subsection{Comparing completeness and contamination}\label{app:comp}
The completeness analysis of \citet{colbert2013} differs in three 
ways from the analysis described in Section~\ref{app:wispcorrections}. First, the automatic line detection 
identifies candidate emission lines through the detection of contiguous pixels
above the measured continuum. Many hot pixels and noise spikes were detected 
in this way and needed to be rejected during the visual inspection phase.
In the updated procedure, this step is supplemented by the continuous 
wavelet transform that selects for emission line shape as well as strength 
above the continuum. This addition, more than any other, serves to remove 
the majority of spurious detections that were identified by the original 
algorithm. The new procedure also implements additional quality checks, 
including a cut on very low EW emission lines designed to remove noise spikes
and a higher S/N threshold of S/N$>$2.31 per pixel (compared with the 
$\sqrt(3)=1.73$). The algorithm automatically rejects any emission line 
candidates detected within 5 pixels of the edge of each spectrum, a 
region where the grism sensitivity decreases rapidly and in which many 
spurious lines were identified by the first version.
Finally, the continuum is estimated on a median-filtered 
spectrum rather than with a spline fit as was used in the first version. This 
approach produces a better fit, especially in regions where the continuum 
changes rapidly. Detecting lines as pixels with excess flux above the 
continuum requires a properly-fit continuum.

Second, in the original procedure, the reviewers identified and fit emission 
lines individually for each source. Once they selected an emission feature 
for fitting -- thus assigning a redshift to the source -- they would step through 
the spectrum fitting Gaussians at the wavelengths expected for other 
lines given the assumed redshift. The widths and central wavelengths of these 
Gaussians were not constrained, and so could vary from line to line in the 
same spectrum. The new procedure fits each spectrum simulataneously, including 
the continuum and all emission lines, and each line is constrained to have 
the same FWHM. Additionally, the reviewer has the option of refitting the 
global continuum during the spectral fitting process, an option that can 
provide better emission line fits and more accurate EW measurements.

The two differences discussed so far refer to the line finding and fitting 
procedure, while the third difference relates specifically to the completeness 
analysis from \citet{colbert2013}. During the visual inspection used to 
calculate the completeness, the reviewers inspected a random 
collection of spectra from real and simulated sources. The goal was to 
avoid the bias that can be introduced when reviewers expect all spectra to 
have emission lines. This step allowed \citet{colbert2013} to calculate 
a contamination rate of 8.5\% due to false emission lines. While we have 
not calculated this for the new line finding procedure, we emphasize that 
the criteria for accepting or rejecting emission line candidates as well as 
the data products included in the inspection (direct images, 2D spectral 
stamps, S/N spectra, and 1D spectra) are the same for both procedures.

\begin{figure}
\plotone{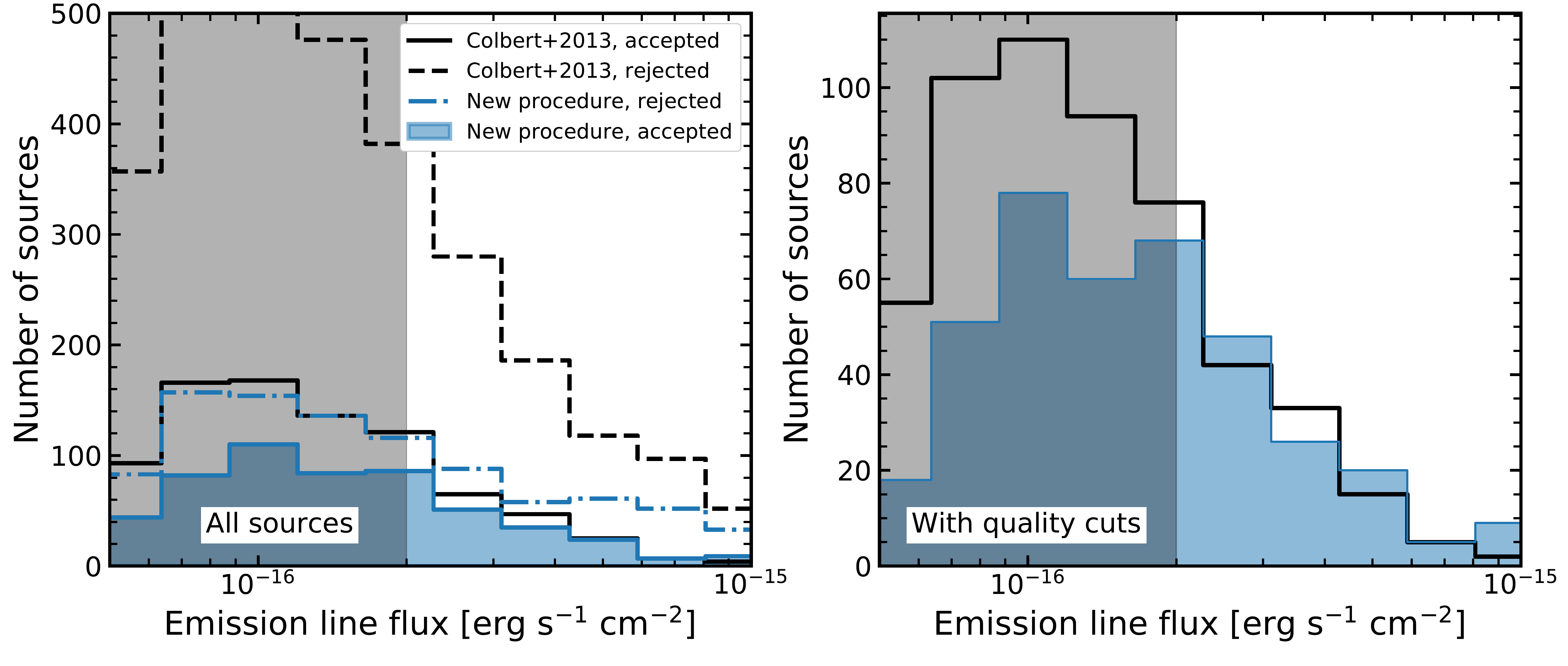}
\caption{
A comparison between the number of emission lines identified as a function of 
line flux by the two versions of the line finding procedure. We include 
the output catalogs from 20 WISP fields that were processed with both versions.
The left panel shows the number of automatically-identified sources that 
were later rejected by reviewers for 
the original (black dashed line) and updated (blue dot-dashed line) versions.
The corresponding distributions of sources that were accepted during visual 
inspection are displayed as a black solid line and a blue filled histogram,
respectively. In the right panel, we show the number of accepted sources 
with quality flags indicating secure redshifts. The automatic detection portion 
of the original procedure identified far more spurious sources, which were 
then removed by reviewers during the visual inspection phase. After visual 
inspection, and especially after applying quality cuts, 
both versions of the procedure identify very similar 
distributions of sources down to the flux limit of the Euclid Wide Survey. 
As final emission line fluxes are measured for accepted sources during 
visual inspection, we note that the emission line fluxes of the rejected 
sources shown here are preliminary estimates produced by the automatic 
detection processes.
\label{fig:linefinding}}
\end{figure}

We have explored the results of the two emission line detection methods 
using 20 WISP fields that have been processed with both versions. In 
Figure~\ref{fig:linefinding}, we show the number of accepted and rejected 
sources as a function of emission line flux. 
In the left panel, the dashed and dot-dashed lines show the number of sources 
identified by the two algorithms that were later rejected by reviewers. 
For a given emission line flux, the first version of the line finding 
algorithm identified $>2\times$ more sources, the majority of which were 
rejected during inspection. Yet the distributions of algorithm-identified and 
reviewer-accepted sources are similar down to the flux limit of the Euclid 
Wide Survey, which is indicated by the grey shaded region. This similarity is 
even greater in the right panel, where we show the same comparison after 
applying quality cuts to the samples that select sources with secure redshifts.
We emphasize here that the y-axes of both panels show the number of sources 
and not a fraction or normalized distribution. 
As we apply the same quality cuts to the WISP and \tdhst\ 
catalogs during the creation of the WS in Section~\ref{sec:selection}, 
the distributions in the right panel are indicative of the performance of the 
two algorithms in our full analysis.

%%%%%%%%%%%%%%%%%%%%%%%%%%%%%%%%%%%%%%%%
\bibliography{bagley_ms}

\end{document}

%% file: definitions.tex
\definecolor{red}{rgb}{1,0,0}
\definecolor{orange}{rgb}{1,0.5,0}
\definecolor{green}{rgb}{0,127,0}
\definecolor{grey}{rgb}{0.6627,0.6627,0.6627}
\definecolor{skyblue}{rgb}{0.53,0.808,0.98}

\newcommand{\eseven}{\times10^{-17}}

\newcommand{\ha}{H$\alpha$}
\newcommand{\hb}{H$\beta$}
\newcommand{\hg}{H$\gamma$}

\newcommand{\oii}{[\ion{O}{2}]}
\newcommand{\sii}{[\ion{S}{2}]}
\newcommand{\siii}{[\ion{S}{3}]}
\newcommand{\oiii}{[\ion{O}{3}]}
\newcommand{\oiiib}{[\ion{O}{3}]~$\lambda5007$}
\newcommand{\hei}{\ion{He}{1}}

\newcommand{\nii}{[\ion{N}{2}]}
\newcommand{\han}{H$\alpha+$[\ion{N}{2}]}

\newcommand{\esc}{erg s$^{-1}$ cm$^{-2}$}

\newcommand{\se}{\textit{Source Extractor}}
\newcommand{\galfit}{\textsc{Galfit}}

\newcommand{\hst}{\textit{HST}}
\newcommand{\rst}{\textit{Roman}}
\newcommand{\tdhst}{3D-HST$+$AGHAST}

\newcommand{\reff}{$R_{\mathrm{eff}}$}